\documentclass[%
 reprint,
superscriptaddress,
nofootinbib,
amsmath,amssymb,
aps,
prstab,
floatfix,
]{revtex4-2}

\usepackage{graphicx}
\usepackage{dcolumn}
\usepackage{bm}
\usepackage[T1]{fontenc}
\usepackage{newtxmath,newtxtext}
\usepackage{mathtools}
\usepackage{physics}
\usepackage{enumitem}
\usepackage[caption=false]{subfig}
\usepackage{derivative}
\usepackage{xr-hyper}
\usepackage{hyperref}
\usepackage{cancel}
\usepackage{xcolor}
\usepackage{makecell}
\usepackage{float}
\usepackage{lipsum}
\usepackage{svg}
\usepackage[caption=false]{subfig}


\usepackage{ragged2e} 
\DeclareCaptionJustification{justified}{\justifying}

\makeatletter
\def\bibsection{%
  \par
  \begingroup
    \baselineskip26\p@
    \bib@device{\hsize}{72\p@}%
  \endgroup
  \nobreak\@nobreaktrue
  \addvspace{19\p@}%
  }%
\makeatother

\begin{document}
\preprint{APS/123-QED}
\title{Incoherent horizontal emittance growth due to the interplay of beam-beam and longitudinal wakefield in crab-waist colliders}



\author{Peter Kicsiny}
\email{peter.kicsiny@cern.ch}
\affiliation{European Organisation for Nuclear Research (CERN), CH 1211 Geneva 23, Switzerland}
\affiliation{\'Ecole Polytechnique F\'ed\'erale de Lausanne (EPFL), Route Cantonale, 1015 Lausanne, Switzerland}

\author{Demin Zhou}
\email{dmzhou@post.kek.jp}
\affiliation{KEK, 1-1 Oho, Tsukuba 305-0801, Japan}
\affiliation{The Graduate University for Advanced Studies, SOKENDAI}

\author{Xavier Buffat}
\affiliation{European Organisation for Nuclear Research (CERN), CH 1211 Geneva 23, Switzerland}

\author{Tatiana Pieloni}
\affiliation{\'Ecole Polytechnique F\'ed\'erale de Lausanne (EPFL), Route Cantonale, 1015 Lausanne, Switzerland}

\author{Mike Seidel}
\affiliation{\'Ecole Polytechnique F\'ed\'erale de Lausanne (EPFL), Route Cantonale, 1015 Lausanne, Switzerland}

\date{\today}

\begin{abstract}
In this paper, we investigate quadrupolar sychrobetatron resonances caused by beam-beam collisions and their interplay with longitudinal wakefields in the context of crab-waist colliders. We present a comprehensive theoretical review of the established theory of sychrobetatron resonances and extend the formalism to explore horizontal sychrobetatron resonances specific to crab-waist colliders. As a case study, we examine incoherent horizontal emittance growth at the SuperKEKB and demonstrate through simulations that the interplay between beam-beam and longitudinal wakefields leads to a horizontal blowup of the bunch size and that the study of the dynamics can be reduced to the horizontal-longitudinal plane, independent of the motion in the vertical dimension. We present extensive simulation results using the codes \texttt{BBWS}, \texttt{PyHEADTAIL} and \texttt{Xsuite}, connect our analytical findings with these findings, and propose strategies to mitigate horizontal blowup.
\end{abstract}

\maketitle


\section{\label{sec:intro}Introduction}

Recent studies suggest that the luminosity performance of SuperKEKB and future circular $e^+e^-$ colliders, all of which will employ the crab-waist scheme with a large Piwinski angle~\cite{Raimondi2006}, can be significantly influenced by beam-beam effects and their interplay with various physics aspects, such as lattice nonlinearities and impedance effects~\cite{Zhou2015IPAC, PhysRevLett.119.134801, kuroo2018cross, zhang2020self, migliorati2021interplay, lin2022coupling, zhang2023combined, zhou2023simulations, ohmi2023beam}. Achieving high luminosities in these colliders critically depends on the crab-waist scheme, which is designed to suppress incoherent beam-beam resonances arising from horizontal crossing and hourglass effects~\cite{pestrikov1993vertical, dikansky2009effect, zobov2010test}. However, while the crab-waist scheme effectively mitigates incoherent resonances, it may not adequately suppress coherent transverse-longitudinal ($x-z$) instabilities that result from collisions at large crossing angles~\cite{PhysRevLett.119.134801, kuroo2018cross}. Collective effects driven by impedance, particularly, alter the dynamics of synchrotron and betatron motions by influencing shifts and spreads of the tunes~\cite{migliorati2018impact}. Recent research highlights that coherent instabilities, driven by the combined effects of beam-beam interactions and impedance, can manifest in both horizontal ($x$)~\cite{zhang2020self, lin2022coupling} and vertical ($y$) planes~\cite{zhang2023combined, ohmi2023beam}. Such combined effects have been observed in earlier studies~\cite{perevedentsev2001simulation, white2014transverse} but gain renewed importance in modern crab-waist collider designs.

The KEKB was an asymmetric electron-positron collider which operated between 1998 and 2010 at KEK in Tsukuba, Japan and achieved a then world record in luminosity ($\sim 2.11\cdot 10^{34}$~cm$^{-2}$s$^{-1}$~\cite{abe2013achievements} at its Belle detector~\cite{ABASHIAN2002117}. An upgrade to the KEKB project, SuperKEKB~\cite{skekb_cdr}, collides a positron beam of 4~GeV in its Low Energy Ring (LER) with an electron beam of 7~GeV in its High Energy Ring (HER). Since its first commissioning in 2016~\cite{Ohnishi:2016yyh, Funakoshi:2016suf}, it has operated successfully through 3 phases and has entered its fourth phase, started in January 2024. Since 2022 it has been the holder of the record luminosity with $4.71\cdot 10^{34}$~cm$^{-2}$s$^{-1}$~\cite{Ohnishi_eeFACT2022, Zhou_2024}. This achievement was made possible by various machine optimizations, such as the application of the crab-waist scheme since 2020~\cite{skekb_cw} based on a similar scheme initially designed for the FCC-ee~\cite{Benedikt:2651299, PhysRevAccelBeams.19.111005, Zobov_2016}, collisions with a large full crossing angle of 83~mrad and with a vertical spot size of the order of hundreds of nanometers.

A significant challenge in crab-waist colliders is the coherent beam-beam-driven $x-z$ instability, which becomes particularly severe near the resonance lines of $2Q_x-kQ_z=\textit{Integer}$~\cite{zhang2023combined}, where $k=2,4,6,\ldots$ with $Q_x$ above half integer. Here, $Q_x$ and $Q_z$ are the horizontal betatron and longitudinal synchrotron tunes, respectively. In~\cite{zhou2023simulations}, a horizontal emittance blowup was observed even when the working point was sufficiently distant from these resonances. This tune-dependent horizontal blowup was observed in both strong-strong beam-beam simulations and machine studies with beams at SuperKEKB. This emittance growth was attributed to synchrobetatron resonances (SBRs) driven by beam-beam interactions, with the resonant conditions altered by beam-beam and impedance induced tune spread. The goal of this paper is to extend previous work by focusing on incoherent horizontal emittance growth driven by the combined effects of beam-beam interactions and longitudinal wakefields. Given the nature of incoherent effects, a weak-strong beam-beam model is sufficient for our analysis.

Our approach builds on methods from earlier works~\cite{PhysRevLett.119.134801, kuroo2018cross, zhang2020self, migliorati2021interplay, lin2022coupling, zhou2023simulations} to investigate the interplay between the longitudinal wakefield and the beam-beam interaction in crab-waist colliders. We begin by revisiting the theory of SBRs driven by beam-beam interactions, a topic with a rich history dating back to the 1970s~\cite{piwinski1977satellite, tennyson1982dynamics, gerasimov1987synchrobetatron, hirata1995analysis, pestrikov1993vertical, dikansky2009effect, Oide:2713385}. Then the analysis is extended to incorporate perturbations due to longitudinal wakefields and detailed beam-beam simulations using several numerical tools are presented to investigate the horizontal emittance growth at the SuperKEKB LER.

The paper is organized as follows: Sec.~\ref{sec:theory} details the  derivation of the fundamental theory of SBRs driven by beam-beam interactions, extending it to include the effects of longitudinal impedance. In Sec.~\ref{sec:codes}, the numerical tools used in this paper are briefly introduced. In Sec.~\ref{sec:simulations} the numerical modeling of the longitudinal wakefields in \texttt{PyHEADTAIL} is introduced, which is interfaced to \texttt{Xsuite}, and a benchmark of \texttt{Xsuite} in a setup featuring longitudinal wakes is shown. In Sec.~\ref{sec:interplay} the simulation results are presented and discussed, and connected with the analytical formalism, to better understand the mechanism of horizontal emittance growth. In addition, mitigation strategies are proposed to reduce the blowup. Finally, Sec.~\ref{sec:summary} provides a summary of the findings and outlines potential directions for future research.

\section{Theory of horizontal synchrobetatron resonances driven by beam-beam interaction}\label{sec:theory}

We follow the formalism of~\cite{pestrikov1993vertical, dikansky2009effect} to derive the weak-strong theory for horizontal SBRs driven by beam-beam interactions. In crab-waist colliders with flat beams and a large Piwinski angle, vertical particle motions around the interaction point (IP) can be largely disregarded when only the horizontal beam dynamics is of concern. This enables the study of horizontal SBRs using a simplified integrated beam-beam potential.

\subsection{Statement of the problem}

According to the weak-strong model, the strong beam in one ring maintains a predefined spatial distribution around the IP. Considering hourglass effects and the crab-waist transform; the normalized spatial distribution of the strong beam is given by
\begin{equation}
    \rho_\text{CW}=\lambda_0(z_0) \rho_\perp(x_0,y_0,z_0;s_0),
    \label{eq:SpatialDistribution1}
\end{equation}
with the longitudinal Gaussian charge density (considering synchrotron radiation effects and ignore impedance effects)
\begin{equation}
    \lambda_0(z_0)=
    \frac{1}{\sqrt{2\pi}\sigma_{z0}} e^{-\frac{z_0^2}{2\sigma_{z0}^2}},
\end{equation}
and transverse charge density
\begin{equation}
    \rho_\perp(x_0,y_0,z_0;s_0)=
    \frac{e^{-\frac{x_0^2}{2\sigma_{x0}^{*2}}-\frac{y_0^2}{2\sigma_{y0}^{*2}\Gamma^2(x_0,\tau)}}}{2\pi\sigma_{x0}^*\sigma_{y0}^*\Gamma(x_0,\tau)}.
    \label{eq:TransverseDistribution1}
\end{equation}
Here, $\sigma_{u0}^*$ with $u=x,y$ indicates the transverse bunch sizes at the IP and $\sigma_{z0}$ the bunch length. Starting from this point onward, the beam parameters of the strong beam are denoted by the subscript ``0''. $\Gamma(x_0,\tau)$ is a function that indicates the density modifications due to the hourglass effects and the crab-waist transform, i.e.
\begin{equation}
    \Gamma(x_0,\tau)=
    \sqrt{1+\frac{1}{\beta_{y0}^{*2}}\left( h_0\tau+\frac{k_2x_0}{\tan (2\theta_c)} \right)^2},
\end{equation}
with $\tau=s_0+z_0$ the distance to the IP considering the longitudinal relative offset $z_0$, $\beta_{y0}^*$ the vertical beta function at the IP, and $\theta_c$ the half crossing angle. $h_0$ is a Boolean value for turning on or off the hourglass effects. $k_2$ indicates the relative strength for the crab-waist transform (i.e., $k_2=0$ and 1 mean no crab-waist and full crab-waist, respectively). When the hourglass effects are neglected and the crab-waist transform is off (i.e., $k_2$=0 and $h_0$=0), there is $\Gamma(x_0,\tau)$=1 and Eq.~\eqref{eq:SpatialDistribution1} reduces to a regular 3D Gaussian distribution.

A single particle from the weak beam circulates around the other ring, receiving nonlinear kicks from the crab-waist sextupoles and experiencing the beam-beam force exerted by the strong beam around the IP. The Hamiltonian for the weak beam in terms of action-angle variables is
\begin{equation}
    K=Q_xJ_x+Q_zJ_z+V_{bb}\delta(\theta),
    \label{eq:EffectiveHamiltonian}
\end{equation}
with the integrated beam-beam potential given by
\begin{align}
    V_{bb}= &-\frac{N_0r_e(1+\cos(2\theta_c))}{\pi\gamma}
    \int_{-\infty}^\infty ds\lambda_0(z_0) \nonumber \\
    & \iint_{-\infty}^\infty
    \frac{dk_xdk_y}{k_x^2+k_y^2}
    \tilde{\rho}_\perp(\vec{k}_\perp, z_0+s_0)
    e^{-ik_xx_0-ik_yy_0}.
    \label{eq:IntegratedBBpotential1}
\end{align}
Here, $J_{x,z}$ are the horizontal and longitudinal action variables; $\theta=2\pi s/C$ with $s$ the orbit distance and $C$ the circumference of the ring; $N_0$ is the bunch population of the strong beam; $r_e$ is the classical radius of electron; $\gamma$ is the relativistic Lorentz factor of the weak beam. The Dirac delta function $\delta(\theta)$ indicates that the beam-beam kick is lumped at the IP. Applying the Fourier transform to Eq.~\eqref{eq:TransverseDistribution1}, we obtain the transverse spectrum as
\begin{equation}
    \tilde{\rho}_\perp(\vec{k}_\perp,\tau)=
    \frac{1}{\sqrt{1+\kappa_0^2}}
    e^{-\frac{(k_x\sigma_{x0}^*+i\eta_0\tau)^2}{2(1+\kappa_0^2)}-\frac{k_y^2\sigma_{y0}^{*2}}{2}\left( 1+\frac{h_0^2\tau^2}{\beta_{y0}^{*2}} \right)}.
\end{equation}
By definitions, $\eta_0=h_0k_2 \zeta_{x0} k_y^2\sigma_{y0}^{*2}/\beta_{y0}^*$ and $\kappa_0=k_2 \zeta_{x0} k_y\sigma_{y0}^*$, where $\zeta_{x0}=\sigma_{x0}^*/(\beta_{y0}^*\tan(2\theta_c))$. It should be noted that $\kappa_0$ indicates a pure crab-waist effect, while $\eta_0$ signifies a coupling of crab-waist and hourglass effects.

Through rotation transforms, the coordinates in the strong beam are expressed by the coordinates in the weak beam's coordinate system as
\begin{equation}
    x_0=-x\cos(2\theta_c)-(z+s)\sin(2\theta_c),
\end{equation}
\begin{equation}
    z_0+s_0=x\sin(2\theta_c)-(z+s)\cos(2\theta_c),
\end{equation}
$y_0=y$, and $s_0=s$. Here, we adopt right-handed coordinates for both beams, with the longitudinal axes aligned with the directions of beam motion. In terms of action-angle variables, the coordinates of a particle in the weak beam are expressed as
\begin{equation}
    x=\sqrt{2\beta_x^*J_x}\cos\psi_x=I_x\cos\psi_x,
\end{equation}
\begin{equation}
    y=\sqrt{2\beta_y^*J_y}\cos\psi_y=I_y\cos\psi_y,
    \label{eq:ymotion}
\end{equation}
\begin{equation}
    z=\sqrt{2\beta_zJ_z}\cos\psi_z=I_z\cos\psi_z.
    \label{eq:zmotion}
\end{equation}
Since we are concerned with the horizontal motion, the hourglass effects on the vertical motion are tentatively ignored here. With the above definitions of particle coordinates, the beam-beam potential of Eq.~\eqref{eq:IntegratedBBpotential1} can be rewritten as
\begin{align}
    V_{bb}= & -\frac{N_0r_er_0}{\pi\gamma}\iint_{-\infty}^\infty \frac{dt_xdt_y}{t_x^2r_0^2+t_y^2}
    \frac{1}{\sqrt{1+k_2^2\zeta_{x0}^2t_y^2}} e^{-\frac{t_y^2}{2}}  \nonumber \\
    & e^{it_x\overline{u}_x-it_yu_y} \int_{-\infty}^\infty d\tau' \lambda(\tau')
    e^{it_x\tau'} \nonumber \\
    & e^{-\frac{h_0^2\zeta_{x0}^2t_y^2}{2}\tau''^2-\frac{\left(t_x-ih_0k_2\zeta_{x0}^2t_y^2\tau''\right)^2}{2(1+k_2^2\zeta_{x0}^2t_y^2)}},
    \label{eq:IntegratedBBpotential2}
\end{align}
with new dimensionless integral variables of $t_x=k_x\sigma_{x0}^*$, $t_y=k_y\sigma_{y0}^*$, and $\tau'=-z_0\tan\theta_c/\sigma_{x0}^*$. Other quantities are defined as $r_0=\sigma_{y0}^*/\sigma_{x0}^*$, $\overline{u}_x=(x+z\tan\theta_c)/\sigma_{x0}^*$, $u_y=y/\sigma_{y0}^*$, and $\tau''=\tau'+(z-x\tan(2\theta_c))\tan\theta_c/\sigma_{x0}^*$. In addition, $\lambda(\tau')$ follows a Gaussian distribution with a normalized RMS width $\phi_0=\sigma_{z0}/\sigma_{x0}^*\tan\theta_c$, which is the Piwinski angle. The substitution of the variable $s$ by $\tau'$ in Eq.~\eqref{eq:IntegratedBBpotential2} is justified under the condition that the dependence of the particle coordinates $x$ and $z$ on $s$ is negligible during the beam-beam interaction.

Since the function in the integral of Eq.~\eqref{eq:IntegratedBBpotential2} is even with respect to $t_y$, we can make the substitution $t_y = |t_x|r_0\tan\theta$. Consequently, Eq.~\eqref{eq:IntegratedBBpotential2} can be rewritten as
\begin{align}
    V_{bb}= & -\frac{2N_0r_e}{\pi\gamma}\int_{-\infty}^\infty \frac{dt_x}{|t_x|}e^{it_x\overline{u}_x}  \nonumber \\
    & \int_0^{\frac{\pi}{2}}
    \frac{d\theta}{\sqrt{1+k_2^2g_0^2\tan^2\theta}} e^{-\frac{1}{2}t_x^2r_0^2\tan^2\theta-i|t_x|r_0u_y\tan\theta} \nonumber \\
    & \int_{-\infty}^\infty d\tau' \lambda(\tau')
    e^{it_x\tau'} e^{-\frac{h_0^2g_0^2\tan^2\theta}{2}\tau''^2} \nonumber \\
    &e^{-\frac{t_x^2\left(1-ih_0k_2\zeta_{x0}^2r_0^2 t_x \tau''\tan\theta\right)^2}{2(1+k_2^2g_0^2\tan^2\theta)}},
    \label{eq:IntegratedBBpotential3}
\end{align}
with $g_0=\zeta_{x0}t_xr_0=\zeta_{y0}t_x$. With the formulation of Eq.~\eqref{eq:IntegratedBBpotential3}, we can perform an order analysis to examine how the evaluation of $V_{bb}$ depends on the beam parameters. With the existence of $\lambda(\tau')$, the integral over $\tau'$ is determined mainly in the region of $|t_x|\sim 1/\phi_0$. For crab-waist colliders, there are typically $\phi_0\gg 1$ (large Piwinski angle), $r_0\ll 1$ (flat beams), $\zeta_{x0}\lesssim 1$ (hourglass condition), and $\zeta_{y0} \ll 1$. The beam dynamics with $|z|\lesssim 10\sigma_{z0}$ is important in studying beam-beam effects. With these configurations, the relevant quantities fairly satisfy $|g_0|\ll 1$, $|t_x|r_0\ll 1$, $\zeta_{y0}^2|t_x\tau''| \ll 1$, and $|g_0\tau''|\ll 1$. Subsequently, for the study of horizontal SBRs in crab-waist colliders with a large Pwinsiki angle, Eq.~\eqref{eq:IntegratedBBpotential3} can be approximated by
\begin{align}
    V_{bb}\approx & -\frac{2N_0r_e}{\pi\gamma}\int_{-\infty}^\infty \frac{dt_x}{|t_x|}e^{it_x\overline{u}_x}  \nonumber \\
    & \int_0^{\frac{\pi}{2}}
    d\theta e^{-\frac{1}{2}t_x^2r_0^2\tan^2\theta}
    \int_{-\infty}^\infty d\tau' \lambda(\tau')
    e^{it_x\tau'}
    e^{-\frac{t_x^2}{2}},
    \label{eq:IntegratedBBpotential4}
\end{align}
with $y=0$ (assuming vertical motion has negligible effects on the horizontal motion). Indeed, Eq.~\eqref{eq:IntegratedBBpotential4} corresponds to the case where $h_0=k_2=0$ (i.e., without crab-waist and hourglass effects). Consequently, the integration of $\theta$ and $\tau'$ in Eq.~\eqref{eq:IntegratedBBpotential4} can be performed analytically, yielding
\begin{align}
    V_{bb}\approx -\frac{N_0r_e}{\gamma}\int_{-\infty}^\infty \frac{dt_x}{|t_x|} G_y\left( \frac{|t_xr_0|}{\sqrt{2}} \right) e^{it_x\overline{u}_x-\frac{1}{2}t_x^2 \left(1+\phi_0^2 \right)},
    \label{eq:IntegratedBBpotential5}
\end{align}
with $G_y(x)=e^{x^2} \text{Erfc}[x]$ where $\text{Erfc}[x]$ is the complementary error function.

The beam-beam potential is further expanded into a sum of Fourier modes:
\begin{equation}
    V_{bb}\delta(\theta)=\sum_{\vec{m},n}
    V_{m_xm_ym_z}e^{i(m_x\psi_x+m_y\psi_y+m_z\psi_z-n\theta)},
    \label{eq:Vbb1}
\end{equation}
with $\vec{m}=(m_x,m_y,m_z)$ and the amplitude term is calculated by
\begin{align}
    V_{m_xm_ym_z}=& \frac{1}{(2\pi)^4}
    \iiint_0^{2\pi} d\psi_xd\psi_yd\psi_z \nonumber \\
    & V_{bb} e^{-i(m_x\psi_x+m_y\psi_y+m_z\psi_z)}.
    \label{eq:Vmxmymz1}
\end{align}
The amplitude-dependent tune shifts away from resonances can be calculated by
\begin{equation}
    \Delta Q_{ub}(J_x,J_y,J_z)=
    \frac{\partial V_{000}}{\partial J_u},
\end{equation}
with $u=x,y,z$.

In the following, we investigate $V_{m_x0m_z}$ to determine the strengths of horizontal SBRs under different conditions.

\subsection{Horizontal synchrobetatron resonances without longitudinal wakefields}

We assume that the impact of the crab-waist and hourglass effects on horizontal SBRs are negligible. Therefore, we start from Eq.~\eqref{eq:IntegratedBBpotential5} to derive the formulae for $V_{m_x0m_z}$.

Substituting Eq.~\eqref{eq:IntegratedBBpotential5} into Eq.~\eqref{eq:Vmxmymz1}, with $x$ and $z$ replaced by action-angle variables and the Jacobi-Anger expansion
\begin{equation}
    e^{iz\cos\phi}=\sum_{m=-\infty}^\infty
    i^mJ_m(z) e^{im\phi},
\end{equation}
the integration over $\psi_{x,z}$ can be done analytically, yielding
\begin{align}
    V_{m_x0m_z}= & -\frac{N_0r_e}{2\pi\gamma} i^{m_x+m_z}
    \int_{-\infty}^{\infty} \frac{dk}{|k|}
    e^{-\frac{k^2}{2}} \nonumber \\
    & G_y
    J_{m_x}(kA_xr_x) J_{m_z}(kA_zr_z),
    \label{eq:Vmx0mz2}
\end{align}
with the integration variable $k=t_x\sqrt{1+\phi_0^2}$. $J_n(x)$ is the $n$-th order Bessel function of the first kind. Here, we define $A_x=I_x/\sigma_{x0}^*$ and $A_z=I_z/\sigma_{z0}$, representing the normalized amplitudes of the betatron and synchrotron motions, respectively. The factors $r_x=1/\sqrt{1+\phi_0^2}$ and $r_z=\phi_0/\sqrt{1+\phi_0^2}$ are scaling factors defined in terms of the Piwinski angle.

The so-called beam-beam strength parameters (i.e., the beam-beam tune shifts at zero amplitudes for betatron and synchrotron motions) can be easily calculated as
\begin{equation}
    \xi_{x0}=\Delta Q_{xb}(0,0,0)=\frac{N_0r_e\beta_x^*}{2\pi\gamma\overline{\sigma}_{x0}(\overline{\sigma}_{x0}+\sigma_{y0}^*)},
    \label{eq:xi_x0}
\end{equation}
\begin{equation}
    \xi_{z0}=\Delta Q_{zb}(0,0,0)=\frac{N_0r_e\beta_z\tan^2\theta_c}{2\pi\gamma\overline{\sigma}_{x0}(\overline{\sigma}_{x0}+\sigma_{y0}^*)},
\end{equation}
with $\overline{\sigma}_{x0}=\sqrt{\sigma_{x0}^{*2}+\sigma_{z0}^2\tan^2\theta_c}$.

\begin{figure}[!ht]	
    \centering
    \includegraphics[width=\columnwidth]{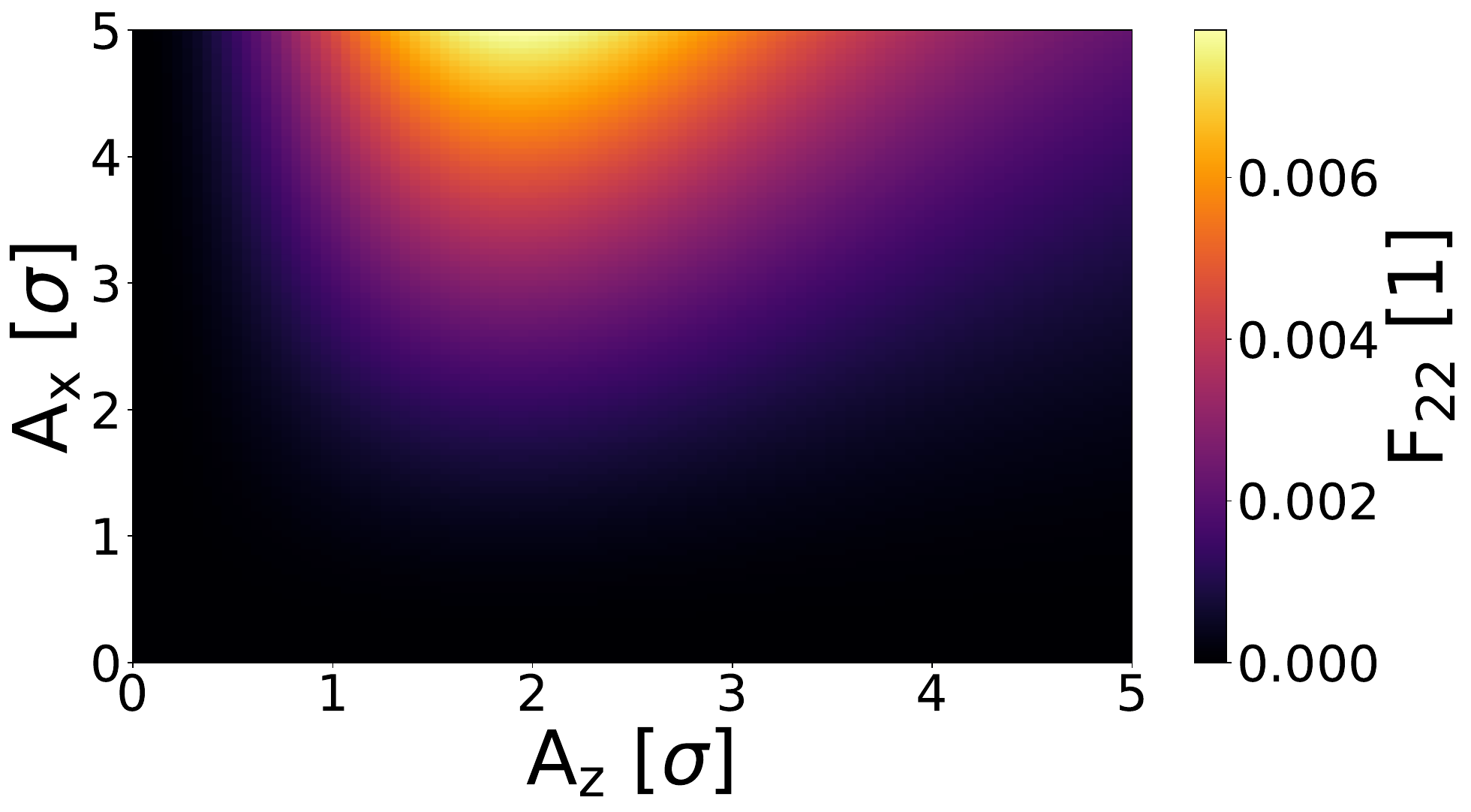}
    \caption{Dimensionless integral $F_{22}$ from Eq.~\eqref{eq:Fmx0mz4} as a function of the normalized amplitudes $A_x$ and $A_z$. The dependence on $A_z$ is approximately quadratic according to Eq.~\eqref{eq:F2mz1}.}
    \label{fig:f22}
\end{figure}

Equation~\eqref{eq:Vmx0mz2} indicates that, without hourglass effects and the crab-waist transformation, only modes with $ m_x + m_z $ being even do not vanish for horizontal SBRs. For these modes of $ m_x + m_z = \textit{even}$, Eq.~\eqref{eq:Vmx0mz2} can be rewritten as
\begin{align}
    V_{m_x0m_z}= & \frac{N_0r_e}{\pi\gamma}
    \int_{0}^{\infty} \frac{dk}{k}
    e^{-\frac{k^2}{2}} \nonumber \\
    & G_y
    J_{m_x}(kA_xr_x) J_{m_z}(kA_zr_z),
    \label{eq:Vmx0mz3}
\end{align}
With flat colliding beams, we can further take $G_y=1$ in the limit of $\sigma_{y0}^*\rightarrow 0$ and approximate the above equation as~\cite{pestrikov1993vertical}
\begin{align}
    V_{m_x0m_z}(A_x,A_z)\approx \frac{N_0r_e}{\pi\gamma} F_{m_xm_z}(A_x,A_z),
    \label{eq:Vmx0mz4}
\end{align}
with
\begin{align}
    F_{m_xm_z}(A_x,A_z) = &
    \int_0^{\infty} \frac{dk}{k}
    e^{-\frac{k^2}{2}} \nonumber \\
    & J_{m_x}(kA_xr_x) J_{m_z}(kA_zr_z).
    \label{eq:Fmx0mz4}
\end{align}
It should be noted that the property of $ \phi_0 \gg 1 $ fundamentally characterizes the features of horizontal SBRs in crab-waist colliders. From it we have $ r_x \approx 1/\phi_0 \ll 1 $ and $ r_z \approx 1 $. The working points of the crab-waist colliders are typically close to half-integer values in the tune space of $(Q_x,Q_y)$. As a result, the resonances $V_{m_x0m_z}$ with $m_x=2$ and $m_z=2,4,6,\ldots$ become particularly significant. Consequently, $F_{2m_z} (A_x,A_z)$ can be approximated by
\begin{align}
    F_{2m_z} \approx 
    \frac{\sqrt{2\pi}}{32\phi_0^2} A_x^2 A_z
    e^{-\frac{A_z^2}{4}} \left[ I_{\frac{m_z-1}{2}} \left( \frac{A_z^2}{4} \right) - I_{\frac{m_z+1}{2}} \left( \frac{A_z^2}{4} \right) \right],
    \label{eq:F2mz1}
\end{align}
where $I_n(x)$ indicates the modified $n$-th order Bessel function of the first kind. Assuming $\phi_0=10$, the contour plot of the dimensionless function $F_{22}(A_x,A_z)$ from Eq.~\eqref{eq:Fmx0mz4} is shown in Fig.~\ref{fig:f22}. It is evident that the amplitude of $F_{22}$ follows the scaling law given by Eq.~\eqref{eq:F2mz1} with respect to both $A_x$ and $A_z$.

\subsection{Horizontal synchrobetatron resonances with longitudinal wakefields}

The longitudinal wakefields can significantly alter synchrotron motion through potential well distortion, thereby interacting with beam-beam interactions and substantially impacting the horizontal SBRs under discussion. Below the microwave instability threshold, the beam maintains a stationary distribution in the longitudinal phase space. The Hamiltonian that describes the longitudinal motion is
\begin{equation}
    H=-\frac{\eta}{2}\delta^2-\frac{\omega_s^2}{2\eta c^2}z^2-I_N\int_z^\infty V_\parallel(z')dz',
    \label{eq:HL}
\end{equation}
with $I_N=Ne^2/(EC)$. Here, $c$ is the speed of light, $e$ is the electron charge and $E$ is the nominal beam energy. Parameters relevant to the weak beam are defined as follows: $\eta$ is the slip factor, $\omega_s$ is the synchrotron frequency and $N$ is the bunch population. The longitudinal wake potential is given by
\begin{equation}
    V_\parallel(z) = \int_{-\infty}^\infty W_\parallel(z-z')\lambda_h(z')dz',
    \label{eq:VL}
\end{equation}
with $\lambda_h(z)$ the Haissinski solution~\cite{haissinski1973exact} and $W_\parallel(z)$ the longitudinal wake function (which characterizes the longitudinal wakefield) for the whole ring. In terms of the longitudinal impedance 
\begin{equation}
    Z_\parallel(k)=\frac{1}{c}\int_{-\infty}^\infty W_\parallel(z)e^{-ikz}dz,
    \label{eq:longimp}
\end{equation}
Eq.~\eqref{eq:VL} is written as
\begin{equation}
    V_\parallel(z)=\frac{c}{2\pi} \int_{-\infty}^\infty  Z_\parallel(k)\tilde{\lambda}_h(k)e^{ikz} dk,
    \label{eq:VLbyZ}
\end{equation}
with $\tilde{\lambda}_h(k)=\int_{-\infty}^\infty \lambda_h(z)e^{-ikz}dz$.

The longitudinal equations of motion for a particle in the weak beam are
\begin{equation}
    \frac{dz}{ds}=-\eta \delta, \quad \frac{d \delta}{ds}=
    \frac{\omega_s^2}{\eta c^2}z - I_N V_\parallel(z).
    \label{eq:Equation_of_motion_delta}
\end{equation}
The stationary Haissinski solution $\lambda_h(z)$ satisfies
\begin{equation}
    \frac{d\lambda_h(z)}{dz}+
    \left[ \frac{z}{\sigma_{z0}^2}-\frac{I_N}{\eta\sigma_\delta^2} V_\parallel(z) \right]
    \lambda_h(z)
    =0.
    \label{eq:de_Haissinski}
\end{equation}
A canonical transformation from ($z, \delta$) to ($J_z, \psi_z$) can be applied to make the Hamiltonian in Eq.~\eqref{eq:HL} phase-independent. This approach was detailed in~\cite{lin2022coupling} to study the combined coherent effects of beam-beam interactions and longitudinal impedance. Here, we aim to develop a simple method to capture the key physics of impedance driven incoherent effects on horizontal SBRs.

For particles with large $J_z$, the synchrotron motion is minimally affected by the wakefields, provided there are no strong sources of resonant impedance in the ring. In this scenario, the theory of SBRs with pure beam-beam interaction, as developed in the previous subsection, is applicable.

\begin{figure}[htb]
   \centering
    \vspace{-1mm}
   \includegraphics*[width=80mm]{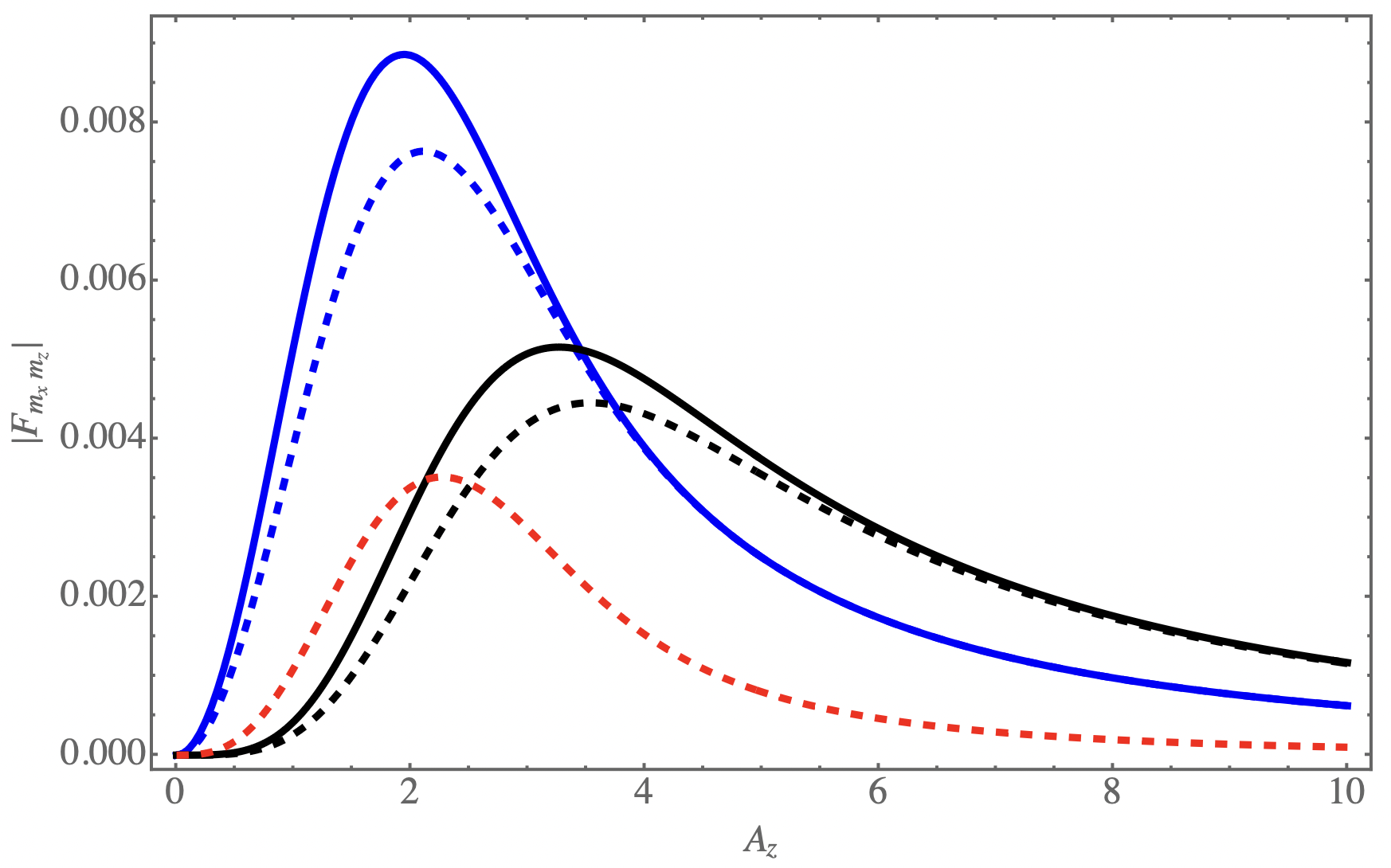}
    \vspace{0mm}
   \caption{Dependence of $F_{2m_z}$ and $F_{2m_z}^w$ on the longitudinal normalized amplitudes, with $\phi_0=10$, $A_x=5$ and $z_m/\sigma_{z0}=0.4$. The solid and dashed lines represent Eqs.~\eqref{eq:Fmx0mz4} and~\eqref{eq:Fmx0mz5}, respectively. Blue and black lines correspond to $m_z=2$ and $4$, while the red dashed line represents $m_z=3$ for $F_{2m_z}^w$.}
   \label{fig:F2mz}
\end{figure}

For particles with small $J_z$, the wakefields can induce a tune shift comparable to the nominal synchrotron tune. Assuming that the Haissinski solution $\lambda_h(z)$ is known with only one peak position at $z_m$ for a given bunch current, we can tentatively expand Eq.~\eqref{eq:VLbyZ} around $z_m$ and keep only the linear term. Consequently, we obtain the second-order differential equation of longitudinal coordinate as
\begin{equation}
    \frac{d^2z}{ds^2}=\frac{\omega_s'^2}{c^2}(z-z_m),
    \label{eq:zmotion2}
\end{equation}
with
\begin{equation}
    \frac{\omega_s'^2}{c^2}=\frac{\omega_s^2}{c^2}
    -\frac{\eta cI_N}{2\pi}
    \int_{-\infty}^\infty ikZ_\parallel(k) \tilde{\lambda}(k)
    e^{ikz_m} dk.
\end{equation}
The peak position can be determined from~\cite{zhou2024theories} with $d\lambda_h(z)/dz|_{z=z_m}=0$:
\begin{equation}
    z_m=\frac{\sigma_{z0}^2I_N}{\eta\sigma_\delta^2} V_\parallel(z_m), 
\end{equation}
assuming that the ring impedance and Haissinski solution are known. The solution of Eq.~\eqref{eq:zmotion2} can be expressed similarly to Eq.~\eqref{eq:zmotion}, but with a shift:
\begin{equation}
    z=z_m + \sqrt{2\beta_zJ_z}\cos\psi_z.
\end{equation}
With this formulation, the amplitude of horizontal SBRs takes the same form as Eq.~\eqref{eq:Vmx0mz4}, but with $F_{m_xm_z}$ replaced by
\begin{align}
    F_{m_xm_z}^w(A_x,A_z) = &
    \frac{1}{2} \int_{-\infty}^{\infty} \frac{dk}{|k|}
    e^{-\frac{k^2}{2}+ik\frac{r_zz_m}{\sigma_{z0}}} \nonumber \\
    & J_{m_x}(kA_xr_x) J_{m_z}(kA_zr_z).
    \label{eq:Fmx0mz5}
\end{align}
The above equation suggests that modes with $m_x+m_z=\textit{odd}$ do not vanish due to the phase term involving $z_m$. This behavior contrasts with the case where there is no wakefield, as predicted by Eq.~\eqref{eq:Vmx0mz2}. It is important to note that Eq.~\eqref{eq:Fmx0mz5} is only valid for small $A_z$, since this condition is required by Eq.~\eqref{eq:zmotion2}. Consider $\phi_0=10$, $A_x=5$ and $z_m/\sigma_{z0}=0.4$ (this number is chosen by referring to Fig.~9 of~\cite{zhou2024theories} for SuperKEKB LER). Equations~\eqref{eq:Fmx0mz4} and~\eqref{eq:Fmx0mz5} with $m_x=2$ and $m_z=2,3,4$ are plotted in Fig.~\ref{fig:F2mz}, illustrating that a nonzero $z_m$ weakens the strengths of modes with $m_z=2,4,\ldots$, while it excites modes with $m_z=3,5,\ldots$.

\subsection{Detuning and resonant conditions}

The particle motion of the weak beam is significantly affected near the resonances described by~\cite{chao2022special}
\begin{equation}
    m_xQ_x + m_yQ_y + m_zQ_z=Integer.
    \label{eq:resonancexyz}
\end{equation}
Here, the tunes $Q_{x,y,z}$ should be considered amplitude-dependent~\cite{pestrikov1993vertical}. For the horizontal SBRs in the presence of beam-beam and wakefield effects, the resonances are described by
\begin{equation}
    m_xQ_x + m_zQ_z=Integer,
    \label{eq:resonancexz}
\end{equation}
with the amplitude dependence of tunes formulated as
\begin{equation}
    Q_x=Q_{x0}+\Delta Q_{xb}(J_x,0,J_z),
\end{equation}
\begin{equation}
    Q_z=Q_{z0}+\Delta Q_{zb}(J_x,0,J_z)+\Delta Q_{zw}(J_z).
    \label{eq:QzJxJz}
\end{equation}
Here, $Q_{x0}$ and $Q_{z0}$ represent the unperturbed tunes at zero-amplitude, and $\Delta Q_{zw}$ represents the longitudinal wakefield effects with
\begin{equation}
    \Delta Q_{zw}(0)=\frac{\omega_s'}{\omega_0}-Q_{z0}.
\end{equation}
Note that with wakefield effects, $J_z=0$ corresponds to $z=z_m$, rather than $z=0$, as in the case without wakefield. The exact formulation of $\Delta Q_{zw}(J_z)$ cannot be derived analytically and requires a numerical solution of the Haissinski equation and further computational efforts, as demonstrated in~\cite{lin2022coupling}.

\section{\label{sec:codes}Beam-beam simulation codes}
Beam-beam simulations for SuperKEKB have been performed since the design stage, using in-house developed macroparticle tracking codes, such as \texttt{BBWS}~\cite{ohmi2003study}, \texttt{BBSS}~\cite{ohmi2000simulation, ohmi2004beam} and \texttt{SAD}~\cite{sad}. It is for the first time that beam-beam studies in the SuperKEKB context are performed using an independent code, \texttt{Xsuite}~\cite{iadarola2023xsuite}. On the one hand, our results demonstrate the power of this tool to simulate the interplay of wakefields and beam-beam interactions. On the other hand, they provide an independent support for previous findings, obtained with the other codes. 

The above mentioned codes are all based on tracking a distribution of particles through a sequence of elements, typically a sequence of magnets (as in \texttt{SAD}), or a linear transfer matrix representing a machine arc segment (as in the others). The beam-beam collision is modeled using the soft-Gaussian approximation~\cite{Bassetti:122227}, which assumes a Gaussian transverse beam profile of the colliding bunches. This approximation is in general applicable for $e^+e^-$ colliders. The beam-beam force is 6D, and is implemented based on the synchrobeam kick algorithm~\cite{Hirata:243013}. Synchrotron radiation in the arcs is modeled in \texttt{SAD} using the full quantum process. In all other codes presented here, an effective model is used which applies exponential damping and Gaussian noise excitation on the dynamical variables, except on $z$. In addition, in all codes there are special elements responsible for other effects such as impedances, space charge, etc.. The impedance modeling in \texttt{Xsuite} is based on another code, \texttt{PyHEADTAIL}~\cite{Oeftiger:2672381}, which is mainly used for coherent beam stability studies at CERN.

\section{\label{sec:simulations}Simulations with longitudinal wakefields}

\subsection{Longitudinal wakefields}

We gradually built up our simulations by adding the different components one by one. In the first iteration, we wanted to make sure that \texttt{Xsuite} can handle wakefields. The wakefield is represented by its own element in the tracking simulation and is based on an implementation in \texttt{PyHEADTAIL}. 
The real valued longitudinal wake function $W_\parallel(z)$ in the spatial (=time, since $z=s-ct$) domain, typically in units of V/pC, is given by the inverse Fourier transform of the complex valued longitudinal impedance $Z_\parallel(k)$, typically in units of $\Omega$, by reversing Eq.~\eqref{eq:longimp}:
\begin{equation}
    W_\parallel(z)=\frac{c}{2 \pi} \int_{-\infty}^{\infty} d k Z_\parallel(k) e^{ikz}.
\end{equation}
The wake kick is applied on the beam in a single element in our tracking model by lumping all the wakefields to that location. To this end, all the various impedance contributions in the SuperKEKB are summed up and will act on the beam as a single perturbation at every tracking turn. When the tracked beam arrives at this element, the particles in each bunch receive a wake kick, depending on their longitudinal coordinate $z$ within their bunch. Each bunch is sliced longitudinally into $N_{sl}$ slices. The particles in the $i$-th slice, where $i\in[0,N_{sl}-1]$ (counting from the head of the bunch), are affected by the wakefield generated by all preceding slices of the same bunch towards the bunch head. From Eq.~\eqref{eq:Equation_of_motion_delta}, the relative energy $\delta_i$ of each particle contained in longitudinal slice $i$ are updated according to~\cite{Rumolo:702717, kevinwake}:
\begin{equation}
\left.\delta_i\right|_{n+1} = \left.\delta_i\right|_n - \frac{e^2}{E}\sum_{j=0}^{i} N_j W_\parallel\left(-(i-j)\Delta z\right),
\end{equation}
where $n$ is the turn index, $N_j$ denotes the number of real charges in slice $j$ (counting from the bunch head), and $\Delta z$ is the width of each longitudinal slice, assuming uniform binning. The negative sign in the wake function argument indicates that the function is evaluated at a distance behind (i.e. towards the bunch tail) the location where the wakefield was excited by the slice $j$. In our simulations the bunch is sliced into $N_{sl}=2000$ slices between the longitudinal range of [$-10\sigma_z$, $10\sigma_z$]. The wake function is then sampled once per slice and assumed to be constant within the given slice.

\begin{figure}[!ht]	
    \centering
    \includegraphics[width=.8\linewidth]{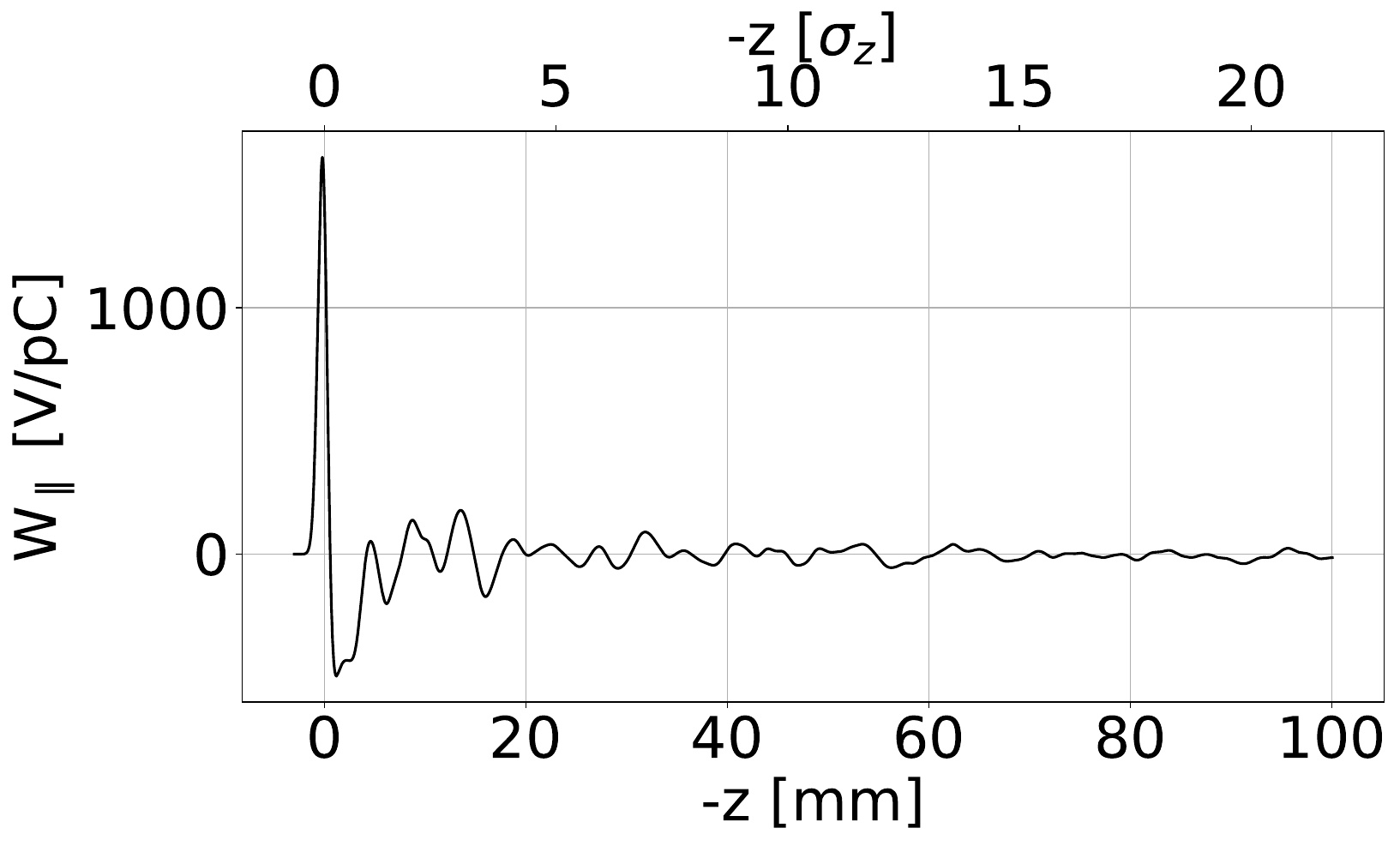}
    \caption{Longitudinal wake function of the SuperKEKB LER used in our simulations. The top axis shows the distance normalized to the RMS bunch length of the LER. Note that $z=0$ indicates the source of the wakefield excitation.}
    \label{fig:wakez}
\end{figure}

The main sources of wakefields in the SuperKEKB LER are RF cavities, resistive walls, collimators, bellows, flanges, clearing electrodes, and tapered beam pipes~\cite{ishibashi2024impedance}. Among these, small-gap collimators are the dominant contributors to inductive impedances. The pseudo-Green function wakes caused by geometric discontinuities were computed using the \texttt{GdfidL}~\cite{gdfidl} and \texttt{ECHO3D}~\cite{echo4d} codes, with a Gaussian driving bunch of $\sigma_z=0.5$~mm. The resistive wall impedance was calculated using the \texttt{IW2D} code~\cite{mounet2010electromagnetic}. The total pseudo-Green function wake, which corresponds to $W_\parallel$ in our notation, was obtained by summing all the calculated wakes~\cite{skekb_wake_Repo}, and it is shown in Fig.~\ref{fig:wakez}. The high-frequency impedance from coherent synchrotron radiation, which does not play a significant role in inducing bunch lengthening or synchrotron tune shifts, is excluded from here.

\subsection{Simulation of microwave instability at SuperKEKB LER}\label{subsec:microwave}

In our first study, we performed a scan of the bunch current in order to reproduce the so-called microwave instability, which is a well-known effect resulting from the presence of longitudinal impedance in the machine~\cite{Zhou:IPAC2014-TUPRI021}.
%
In the bunch current range of  0-4~mA, we tracked 10$^6$ macroparticles. Our model includes a linear transfer matrix representing the full machine arc with effective synchrotron radiation (exponential damping and Gaussian noise excitation), and the longitudinal wakefield element, in this order. The machine parameters used in our simulations are summarized in Table~\ref{tab:params}.

\begin{table}[h!]
\caption{\label{tab:params} SuperKEKB machine parameters used in all simulations in this paper. The values correspond to the machine setup of April 5 2022~\cite{zhou2023simulations}, except for $\varepsilon_y$ of LER. The bunch current $I$ refers to a single bunch and $\theta_c$ is the half crossing angle.}
    \begin{ruledtabular}
    \begin{tabular}{lcc}
  & \textbf{LER} & \textbf{HER}\\
    \hline
                         $C$ [m] & \multicolumn{2}{c}{3016.315} \\
               $\theta_c$ [mrad] & \multicolumn{2}{c}{41.5} \\
                       $E$ [GeV] &      4 & 7.00729 \\
               $N_0$ [$10^{10}$] &  4.459 &   3.579 \\
                        $I$ [mA] &   0.71 &    0.57 \\
          $\alpha_c$ [10$^{-4}$] & 2.9691 &  4.5428 \\
            $\varepsilon_x$ [nm] &    4.0 &     4.6 \\
            $\varepsilon_y$ [pm] &     20 &      35 \\
                $\beta_x^*$ [mm] &     80 &      60 \\
                $\beta_y^*$ [mm] &      1 &       1 \\
         $\sigma_{x}^*$ [$\mu$m] &  17.89 &   16.61 \\
         $\sigma_{y}^*$ [$\mu$m] &   0.14 &    0.19 \\
               $\sigma_{z}$ [mm] &    4.6 &     5.1 \\
   $\sigma_{\delta}$ [$10^{-4}$] &   7.52 &    6.24 \\
          $Q_x$ [1] & 44.524 & 45.532 \\
          $Q_y$ [1] & 46.589 & 43.572 \\
          $Q_z$ [1] &  0.023 &  0.027 \\
        $\xi_x$ [1] & 0.0036 & 0.0024 \\
        $\xi_y$ [1] &  0.056 &  0.044 \\
$\xi_z$ [10$^{-4}$] &   4.78 &   5.58 \\
          $U_{\text{SR}}$ [MeV] &   1.76 &   2.43 \\
    $\tau_{z,\text{SR}}$ [turn] &   2271 &   2881 \\
                     $k_2$ [\%] &     80 &     40 \\
\end{tabular}
\end{ruledtabular}
\end{table}

\begin{figure}[!ht]
    \subfloat[\label{fig:szd_a}Bunch length.]{
	\includegraphics[width=\columnwidth]{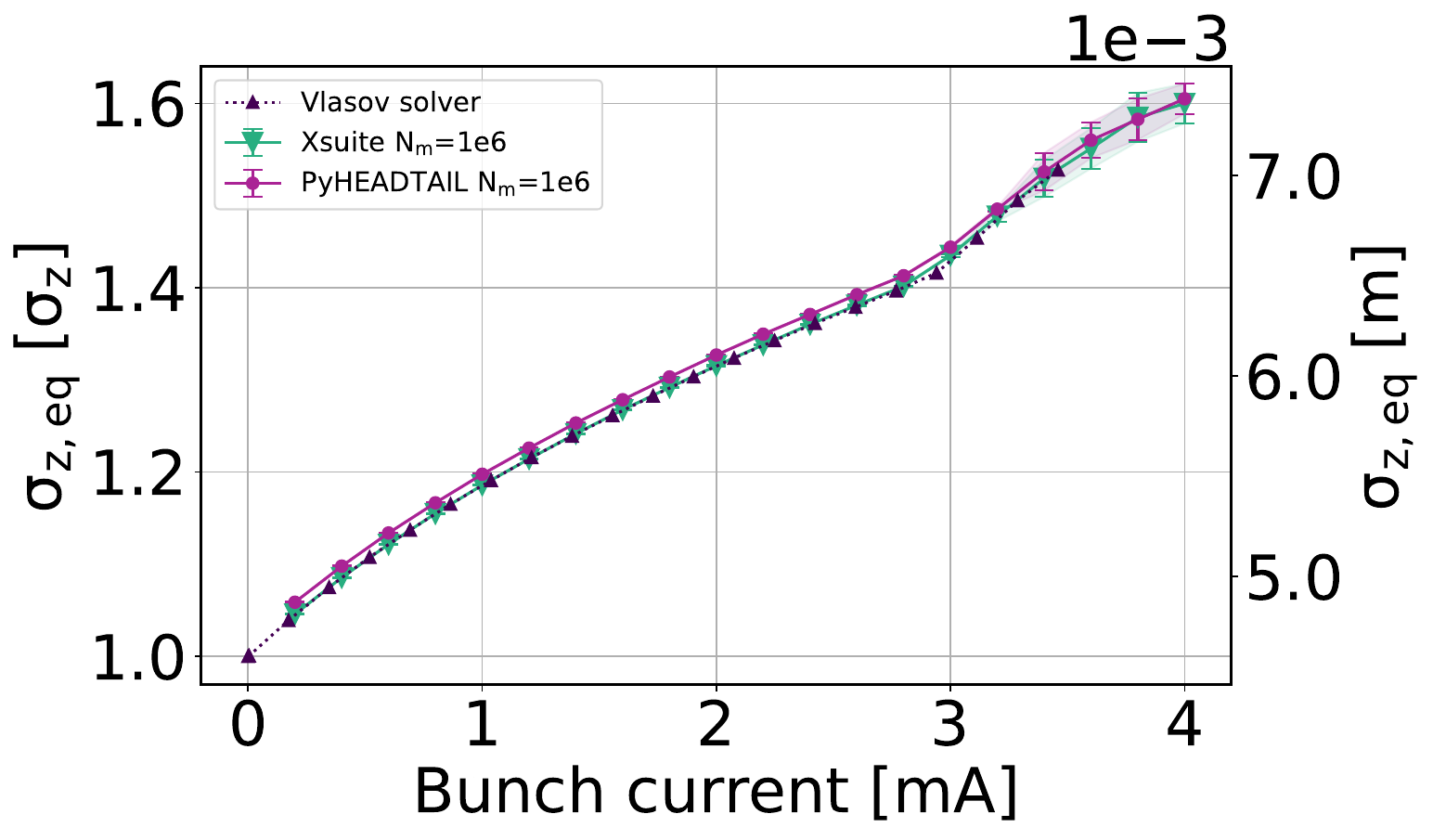}}
	\\	
    \subfloat[\label{fig:szd_b}Energy spread.]{
	\includegraphics[width=\columnwidth]{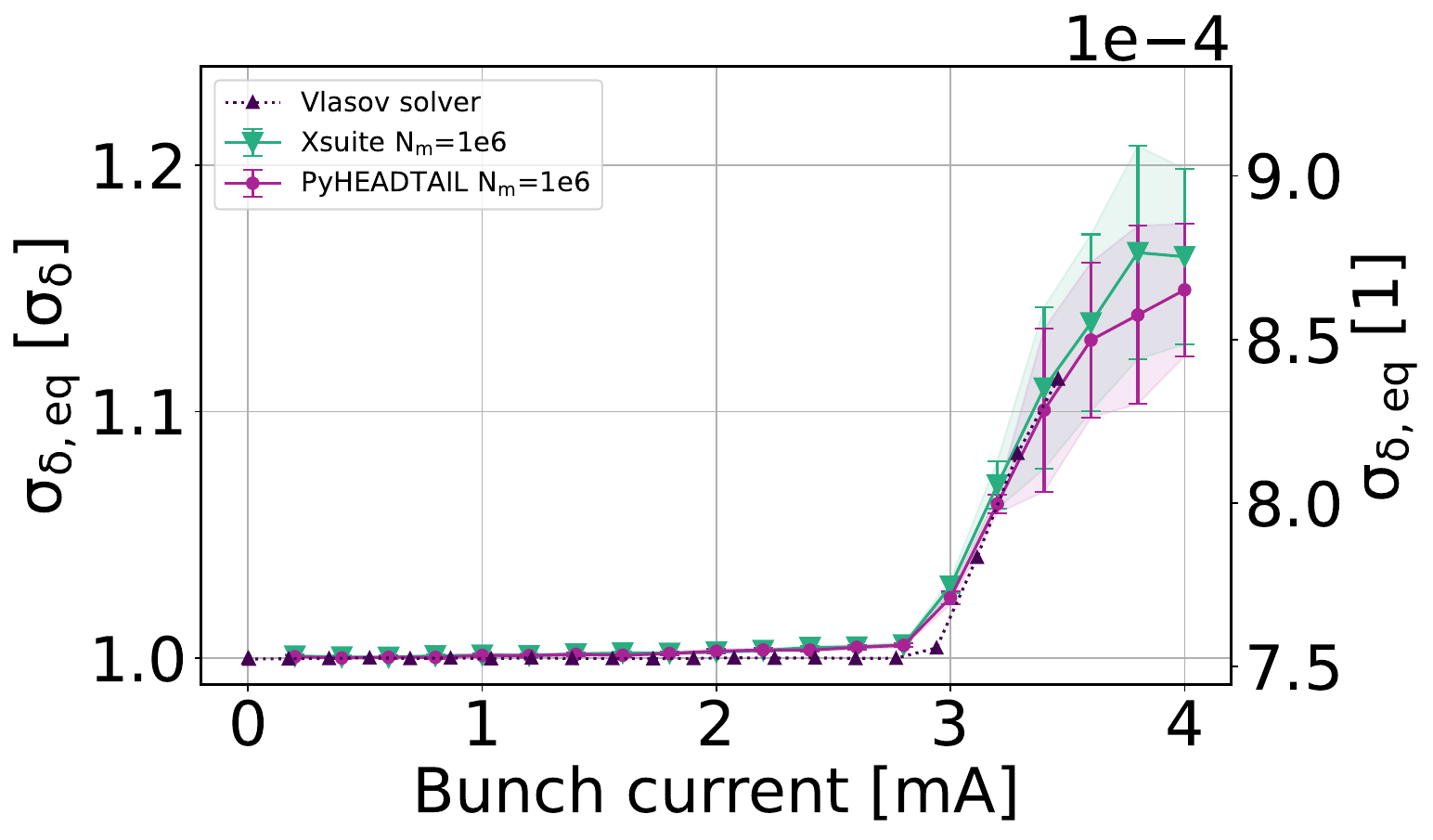}}
    \caption{Equilibrium RMS of bunch length (a) and energy spread (b), calculated from the statistics of the last 5000 turns, as a function of the initial bunch current.}
    \label{fig:szd}
\end{figure}

We have performed the scan with \texttt{Xsuite}, \texttt{PyHEADTAIL}, \texttt{BBWS} and a numerical solver for the Vlasov equation described in~\cite{bane2010threshold}. \texttt{Xsuite} requires an interface to \texttt{PyHEADTAIL} for loading the wakefield data and converting it into a format that can be processed by \texttt{Xsuite}, while the Vlasov solver and \texttt{BBWS} have their own wakefield processing routines. The equilibrium RMS bunch sizes after tracking $2\cdot10^4$ turns with all codes are presented in Fig.~\ref{fig:szd}.

With this benchmark, we have verified that \texttt{Xsuite}, \texttt{PyHEADTAIL} and \texttt{BBWS} reproduce the same physics since the results with these codes are in good agreement with each other. The behavior of the equilibrium energy spread can be divided into 2 regimes, depending on the bunch current. From Fig.~\ref{fig:szd_b} it can be seen that at low bunch currents, i.e. below a given threshold current $I_c$, there is no blowup in the equilibrium energy spread $\sigma_{\delta, \text{eq}}$. Above $I_c$, $\sigma_{\delta, \text{eq}}$ shows a linear dependence on the current, which is caused by the so-called microwave instability. Our simulations demonstrated that with $10^6$ macroparticles the agreement between the tracking codes and the Vlasov solver was highly satisfactory. This study also demonstrated that in the first regime, i.e. at low bunch currents, one can already obtain statistically converged equilibrium dynamics by using $10^5$ macroparticles.

We note that our effective synchrotron radiation model only updates the macroparticle variable $\delta$ instead of both $z$ and $\delta$. We observed that the equilibrium RMS beam sizes and centroids for the longitudinal variables differ when both variables are updated. In pure beam-beam simulations without wakefields, whether $z$ is updated makes no significant difference. However, we argue that in simulations that combine beam-beam interactions and impedance models, updating $z$ in the synchrotron radiation model introduces artificial noise into the bunch's longitudinal coordinate. This is because in these simulations, synchrotron radiation effects, one-turn propagation, and wakefield kicks are modeled as concentrated at a single point, while in reality, both radiation and wakefield effects are distributed around the ring. Updating $\delta$ is justified as a particle that loses energy due to synchrotron radiation will travel a shorter path (assuming that we are above transition), causing it to arrive at the same point on the next turn with a more positive $z$. This effect is naturally accounted for when updating $\delta$, but artificial updating $z$ should be avoided. Nevertheless, from Fig.~\ref{fig:wakez} it can be seen that the wakefield strength changes rapidly at small coordinates close to the source of excitation. Therefore an artificially introduced variation of $z$ will have a significant effect on where the wakefield is sampled and therefore on the beam dynamics. An alternative approach is to employ a different method, e.g., similar to the model used in \texttt{LIFETRAC}~\cite{shatilov2005lifetrac} or the formalism described in~\cite{ohmi1994beam}.

\section{\label{sec:interplay}Interplay of the beam-beam interaction with the longitudinal wakefield}

In the weak-strong beam-beam model, only one beam (the weak one) is tracked, while the other beam (the strong one) is frozen. In the soft-Gaussian approach, the strong beam is represented by the statistical moments of its longitudinal slices. Such a weak-strong model aligns with the analytical theory of SBRs described in Sec.~\ref{sec:theory}. 

\begin{figure}
    \subfloat[\label{fig:qx_scan_ws_x}Horizontal RMS.]{
	\includegraphics[width=\columnwidth]{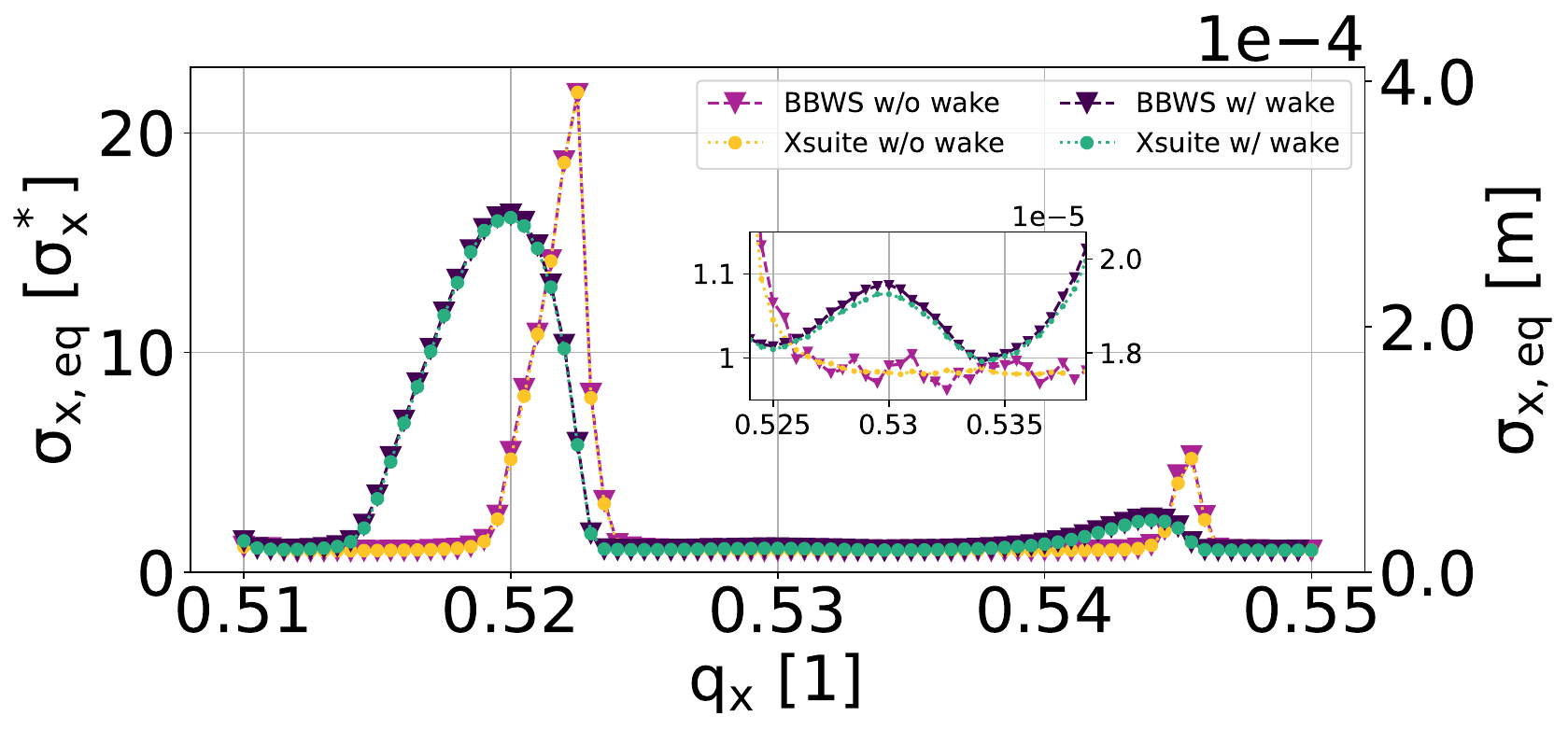}}
	\\	
    \subfloat[\label{fig:qx_scan_ws_y}Vertical RMS.]{
	\includegraphics[width=\columnwidth]{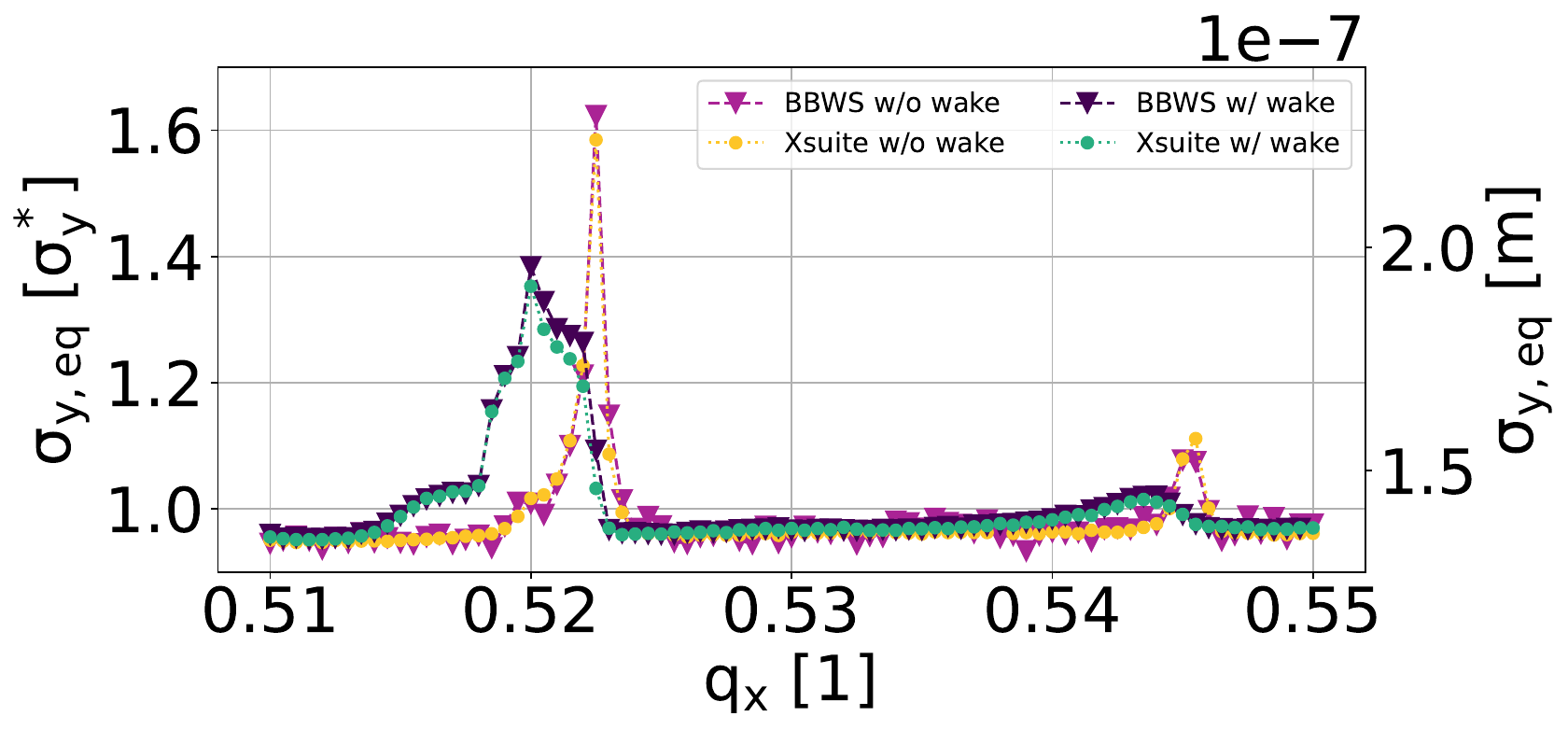}}
	\\	
    \subfloat[\label{fig:qx_scan_ws_z}Longitudinal RMS.]{
        \includegraphics[width=\columnwidth]{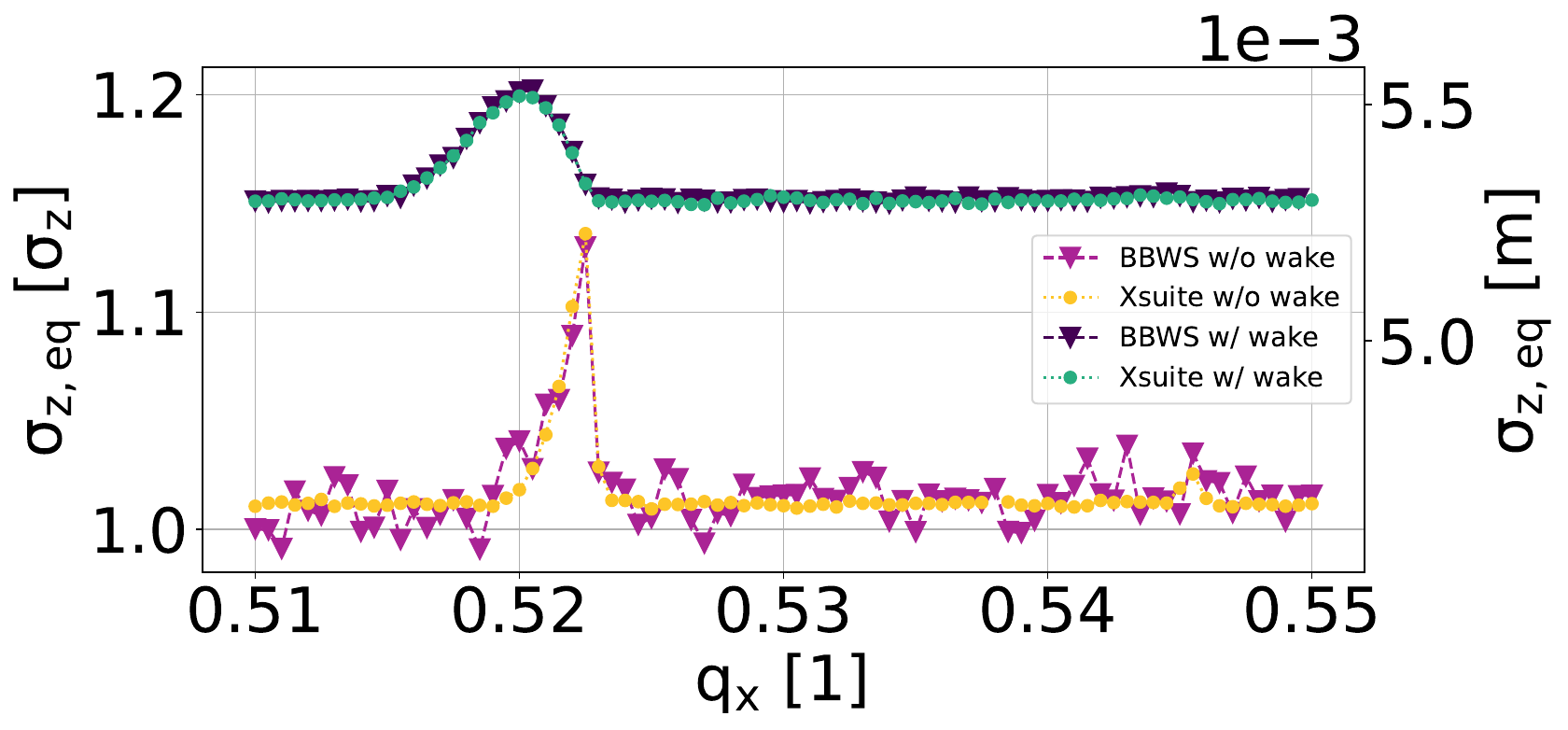}}
    \caption{Equilibrium RMS beam sizes of the LER, calculated from the last 5000 turns as a function of the horizontal fractional tune $q_x$, with and without wakefields, simulated with \texttt{BBWS} and \texttt{Xsuite}. All simulations include a beam-beam collision in the weak-strong model, including the crab-waist scheme.}
    \label{fig:qx_scan_ws}
\end{figure}

\begin{figure}
    \includegraphics[width=\columnwidth]    {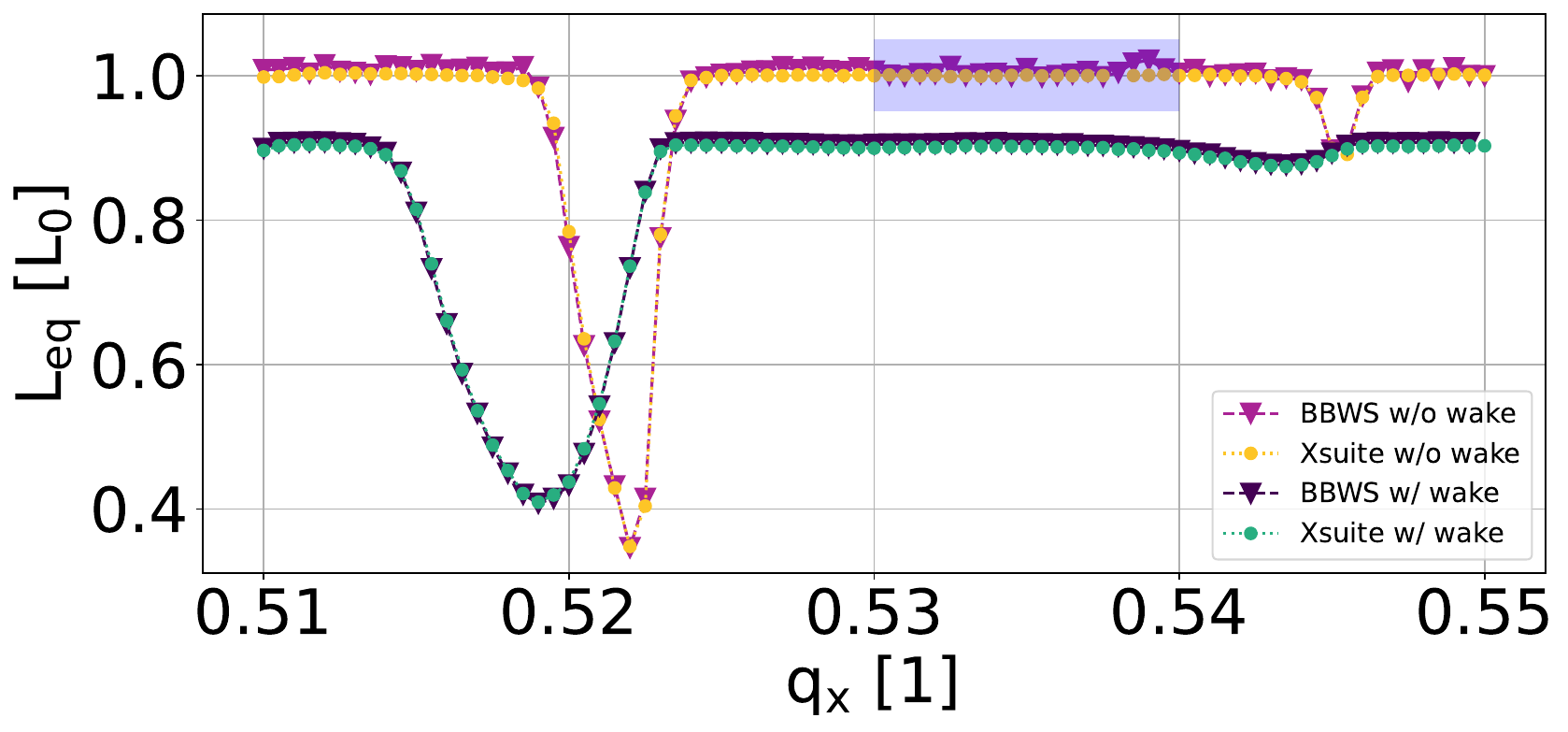}
    \caption{Equilibrium luminosity, calculated from the last 5000 turns as a function of the horizontal fractional tune $q_x$, with and without wakefields, simulated with \texttt{BBWS} and \texttt{Xsuite}. Data are normalized to the mean luminosity 
    $L_0=3.73\times10^{31}$~cm$^{-2}$ s$^{-1}$, obtained from the simulations, in the range $q_x\in[0.53,0.54]$ without wakefield, indicated with a blue rectangle. All simulations include a beam-beam collision in the weak-strong model, including the crab-waist scheme.}
    \label{fig:qx_scan_ws_lumi}
\end{figure}

We performed two sets of simulations: one without and one with the longitudinal wakefield in the tracking model. In both cases our model contains one beam-beam collision per turn, and we track the LER beam with the SuperKEKB parameters~\cite{zhou2023simulations} as shown in Table~\ref{tab:params}. We performed our tracking simulations using the LER beam as the weak one. Because of its lower beam energy, the LER beam is more sensitive to the resonances; however, our results equally apply for the HER too. The tracking model included $2\cdot 10^4$ turns and $10^5$ macroparticles. Using this number of macroparticles yields statistically converged dynamics with the bunch current of the SuperKEKB LER (0.71~mA), which we verified in our study of the microwave instability, presented in Sec.~\ref{subsec:microwave}. We performed each simulation with both \texttt{Xsuite} and \texttt{BBWS}, with and without using the crab-waist scheme. We used the following order of elements in our model: wakefield kick, beam-beam kick, linear arc with synchrotron radiation, observation point. With this setup, we scanned the horizontal fractional tune over the range $q_x\in$ [0.51, 0.55] and with each working point, we recorded the equilibrium beam parameters at the observation point. The equilibrium RMS beam sizes as a function of the fractional tune $q_x$, with crab-waist transformation enabled, are shown in Fig.~\ref{fig:qx_scan_ws}. The equilibrium luminosity, normalized to that with a working point far away from resonances, is shown in Fig.~\ref{fig:qx_scan_ws_lumi}. In the following, we examine these simulation results and highlight their connections to the theories outlined in Sec.~\ref{sec:theory}.


The first observation is that \texttt{Xsuite} and \texttt{BBWS} are in perfect agreement regarding the equilibrium horizontal beam sizes and luminosity, in all the setups considered. There is negligible difference with respect to the vertical and longitudinal equilibrium, which could be attributed to numerics. This benchmark confirms that the transfer maps for the beam-beam interaction and the wakefield kick are correctly implemented in \texttt{Xsuite}. Moreover, we will primarily refer to \texttt{Xsuite} results in the simulations presented in this paper, as it allows for more detailed post-analysis due to its Python-powered flexibility.

As a consequence of beam-beam collisions with a crossing angle, SBR peaks appear at tunes described by Eq.~\eqref{eq:QzJxJz}. At these tunes, the beam size blows up. Due to the tune spread caused by the amplitude dependence of individual particles' betatron and synchrotron tunes, the positions of these peaks are shifted, and their widths are broadened. By turning on the longitudinal wakefield, the peaks shift further towards lower tunes and become even wider. This negative shift is attributed to the amplitude dependent spread of incoherent synchrotron tunes of particles within the bunch due to the potential well distortion. Particles with lower synchrotron amplitudes experience the largest tune shifts as a result of the wakefield kick, leading to this effect. In addition to strong resonances at $(m_x,m_z)=(2,2)$ and $(2,4)$, we also observe a weak horizontal blowup near the resonance of $(m_x,m_z)=(2,3)$ (see Fig.~\ref{fig:qx_scan_ws_x}). This weak blowup reproduces the results of strong-strong beam-beam simulations and has been observed in SuperKEKB with beams, as presented in~\cite{zhou2023simulations}. It only appears in simulations when the longitudinal wakefield kick is activated. Therefore, we conclude that this is an incoherent effect resulting from the combined effect of the beam-beam interaction and the longitudinal wakefield, and it is predicted by Eq.~\eqref{eq:Fmx0mz5}. The shift of the bunch peak in the longitudinal direction caused by the wakefield makes the beam-beam force asymmetric around the center of the bunch, leading to the appearance of resonances with $m_x+m_z=\textit{odd}$. In machine operation, this peak shift can be partially mitigated by adjusting the RF system phase (not the acceleration phase, which is fixed by a balance of energy loss and gain) to adjust the arrival time of the beam at the IP, thus suppressing these resonances. However, due to potential well distortion from the longitudinal wakefield, which causes the bunch profile to tilt along the $z$-axis~\cite{zhou2024theories}, resonances with $m_x+m_z=\textit{odd}$ are expected to persist and cannot be fully eliminated.

\begin{figure}
    \subfloat[\label{fig:qx_scan_cw_on_off_x}Horizontal RMS.]{
	\includegraphics[width=\columnwidth]{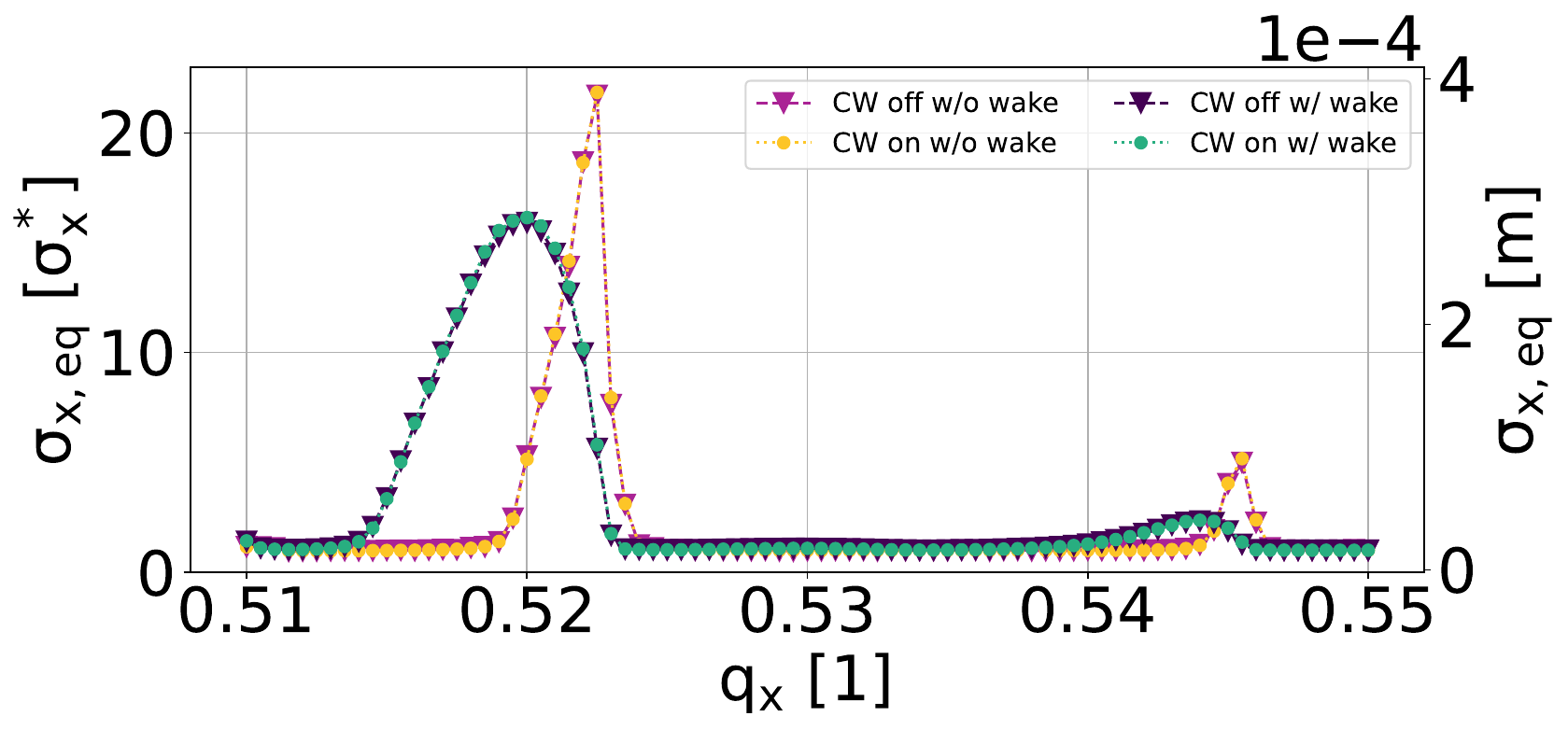}}
	\\	
    \subfloat[\label{fig:qx_scan_cw_on_off_y}Vertical RMS.]{
	\includegraphics[width=\columnwidth]{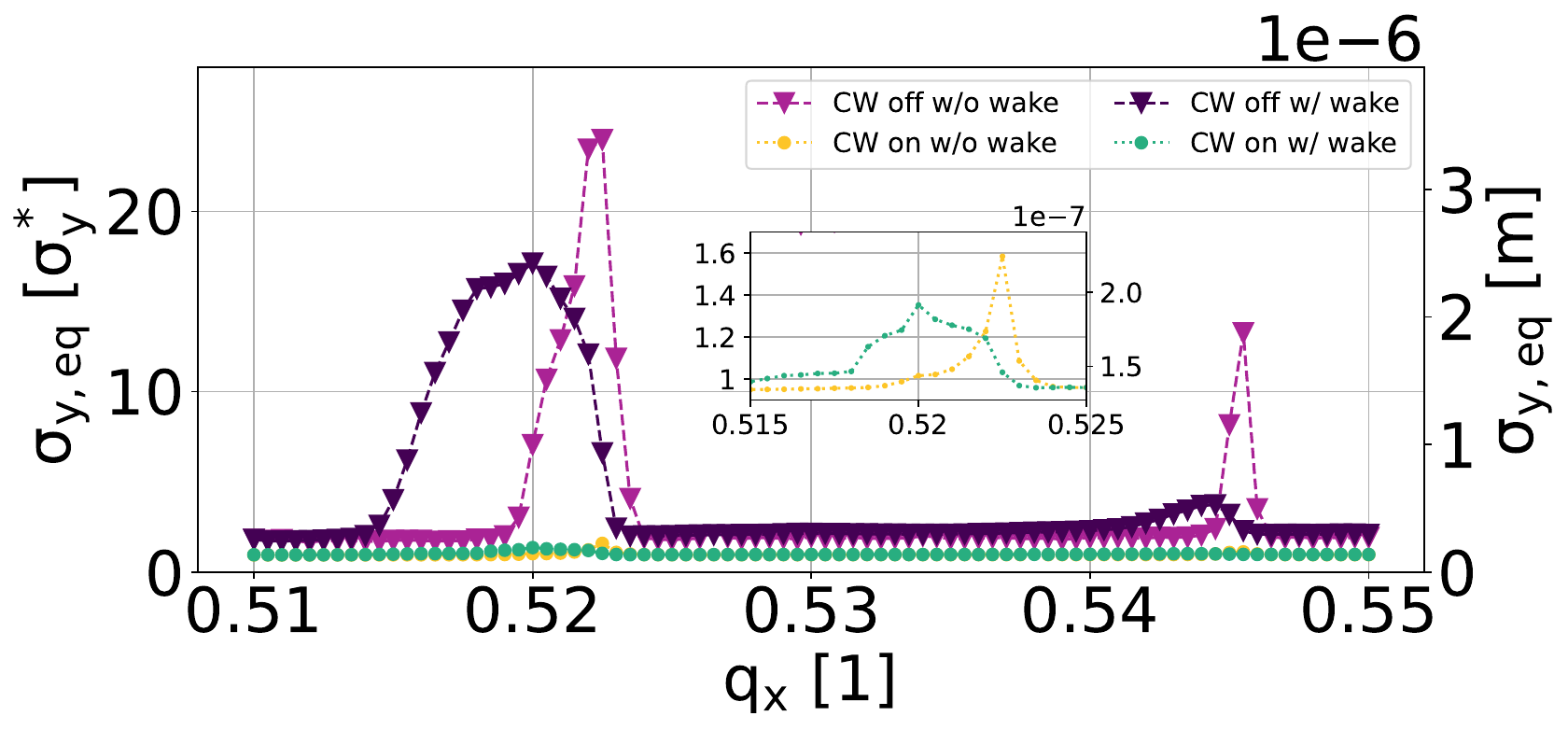}}
    \caption{Equilibrium transverse RMS beam sizes of the LER, calculated from the last 5000 turns as a function of the horizontal fractional tune $q_x$, with and without wakefields, simulated with \texttt{Xsuite}. All simulations include a beam-beam collision in the weak-strong model. CW stands for crab-waist.}
    \label{fig:qx_scan_cw_on_off}
\end{figure}

The crab-waist transform does not visibly affect the shape or location of the peaks in horizontal beam-size blowups described by Eq.~\eqref{eq:resonancexz}, as shown in Fig.~\ref{fig:qx_scan_cw_on_off_x}. This aligns with the order analysis following Eq.~\eqref{eq:IntegratedBBpotential3} to reach Eq.~\eqref{eq:IntegratedBBpotential4}. The main effect of the crab-waist is in the vertical plane, where it reduces vertical blowup caused by beam-beam driven betatron resonances, as shown in Fig.~\ref{fig:qx_scan_cw_on_off_y}. The suppression of vertical blowup by the crab-waist is also discussed in detail in~\cite{zhou2023simulations}. However, when the horizontal blowup becomes severe, especially near SBRs, the crab-waist transform becomes less effective in suppressing the nonlinear $x-y$ couplings caused by beam-beam, leading to a residual vertical blowup (see the zoomed-in subfigure of Fig.~\ref{fig:qx_scan_cw_on_off_y} which replicates Fig.~\ref{fig:qx_scan_ws_y}) and results in a loss of luminosity (see Fig.~\ref{fig:qx_scan_ws_lumi}). Hourglass effects are another contributing factor to the luminosity reduction: when due to the blowup the horizontal bunch size becomes larger, more collisions will take place at larger $\beta_y^*$, due to the large crossing angle.

In the following, we provide a more detailed post-analysis of simulation data to further explore the horizontal beam size growth induced by beam-beam resonances.



\subsection{Investigation of the beam distribution}

\begin{figure}
    \subfloat[\label{fig:xdist_a}$q_x$=0.522.]{
	\includegraphics[width=\columnwidth]{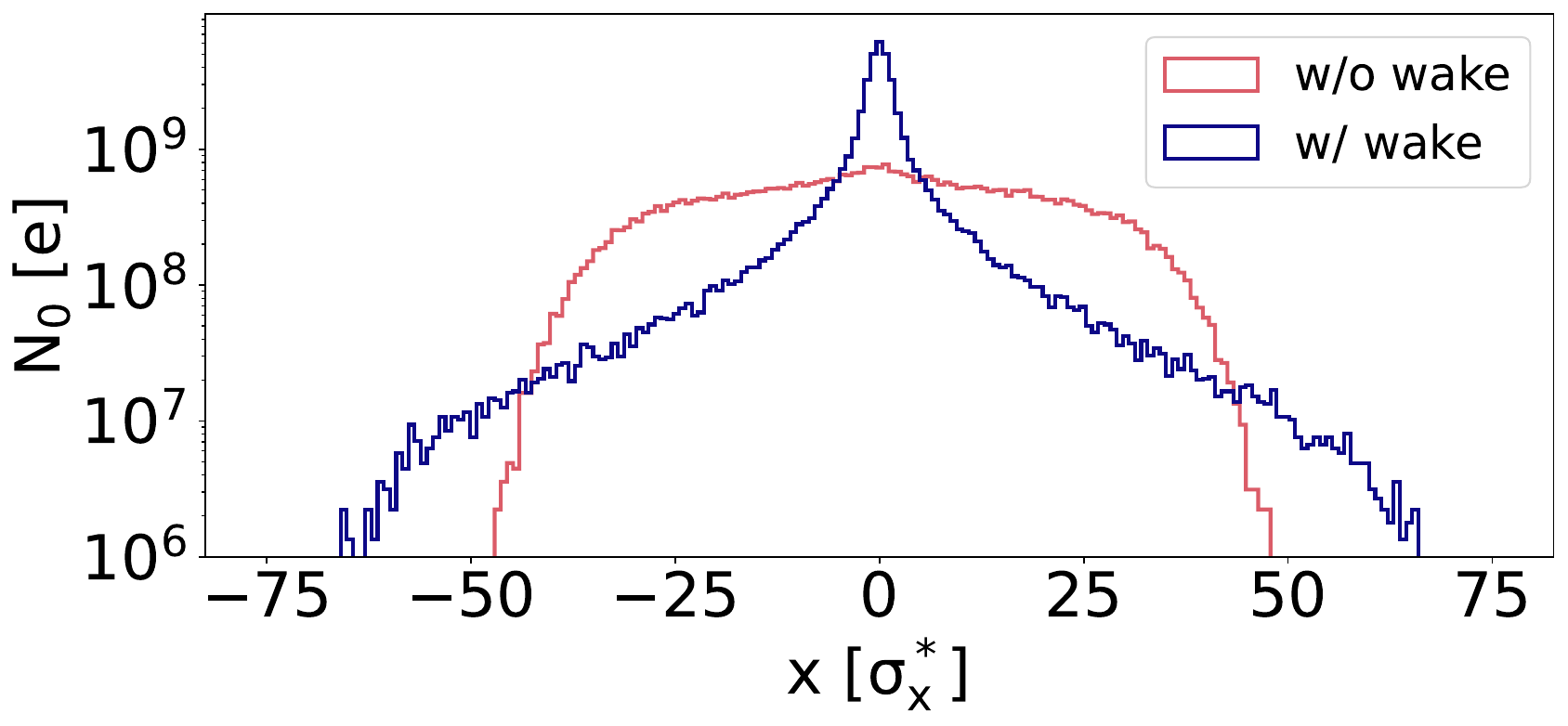}}
	\\	
    \subfloat[\label{fig:xdist_b}$q_x$=0.53.]{
	\includegraphics[width=\columnwidth]{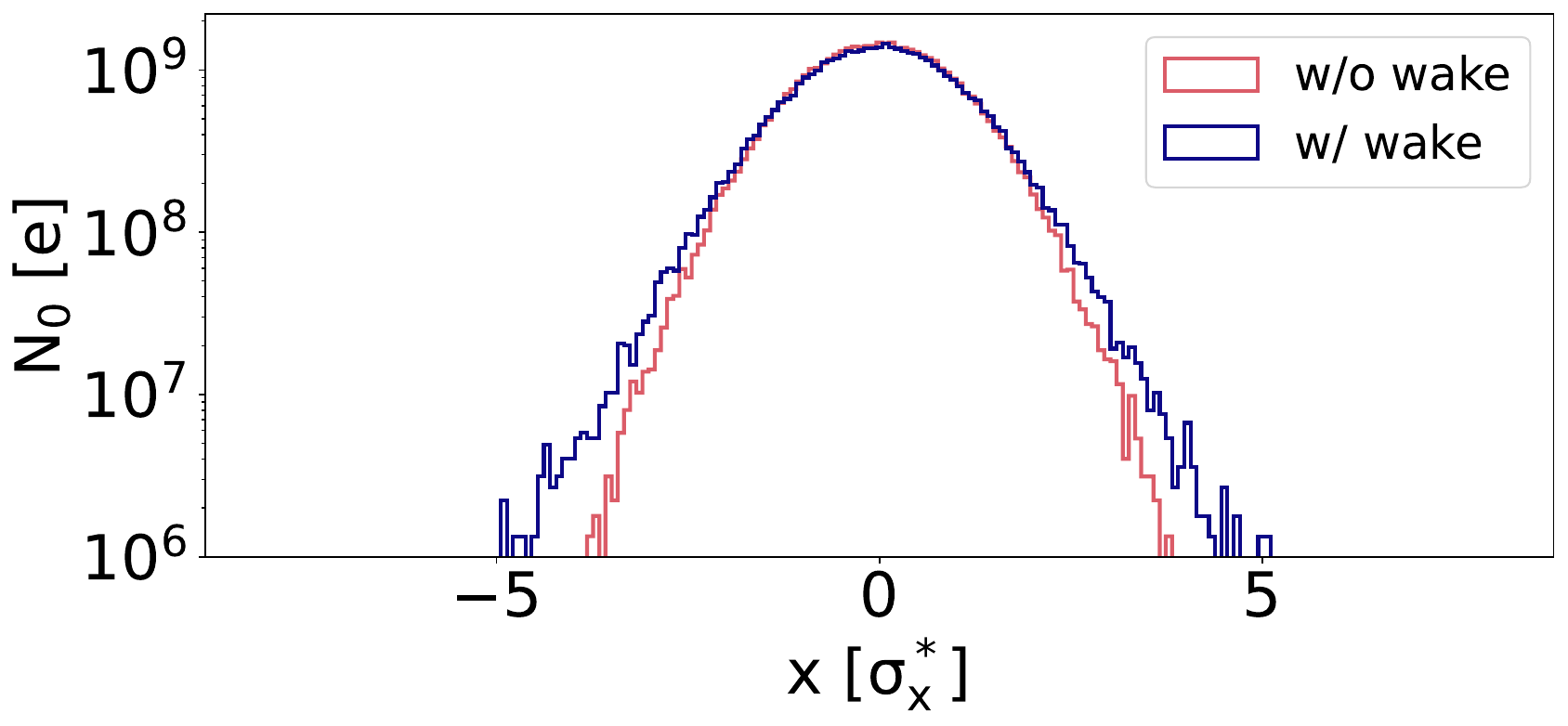}}
    \caption{Horizontal equilibrium bunch distribution with (blue) and without (red) longitudinal wakefields at two selected horizontal fractional tunes. The crab-waist scheme is included in both simulations.}
    \label{fig:xdist}
\end{figure}

Near the SBRs, the equilibrium bunch distribution undergoes significant blowup. Figure~\ref{fig:xdist} illustrates the equilibrium horizontal distribution of charge intensity after tracking for $2\cdot 10^4$ turns, with and without longitudinal wakefields. The distribution is shown for the fractional tune values of $q_x=0.522$, which falls within the resonant region around $2Q_x-2Q_z=\textit{Integer}$, and $q_x=0.53$, which lies outside this region. In the resonant case, without wakefields, the beam shows a maximum blowup, including the core, and the beam tail extends up to $\pm50\sigma_x^*$. With longitudinal wakes present, the tail blowup increases beyond $\pm70\sigma_x^*$. Due to the synchrotron tune spread caused by the longitudinal wakefield, the core blowup is less pronounced than without the wakefield. However, away from this resonance, as depicted in Fig.~\ref{fig:xdist_b}, the wakefield only slightly increases the spread of the equilibrium distribution, corresponding to the excitation of the $2Q_x-3Q_z=\textit{Integer}$ resonance as per Eq.~\eqref{eq:Fmx0mz5}.

Horizontal blowup caused by beam-beam interactions, particularly in the beam tail, is especially detrimental and can be linked to reduced beam lifetime and increased detector background. It also leads to lower injection efficiency as the injected beam will also receive resonant beam-beam kicks from the opposing beam, similar to the stored beam. This issue has already been explored in SuperKEKB~\cite{zhou2023simulations} and it is suggested to be further examined in future studies.

\subsection{Investigation of the amplitude dependence}

\begin{figure}
    \subfloat[\label{fig:sx_a}Without longitudinal wakefield.]{
	\includegraphics[width=\columnwidth]{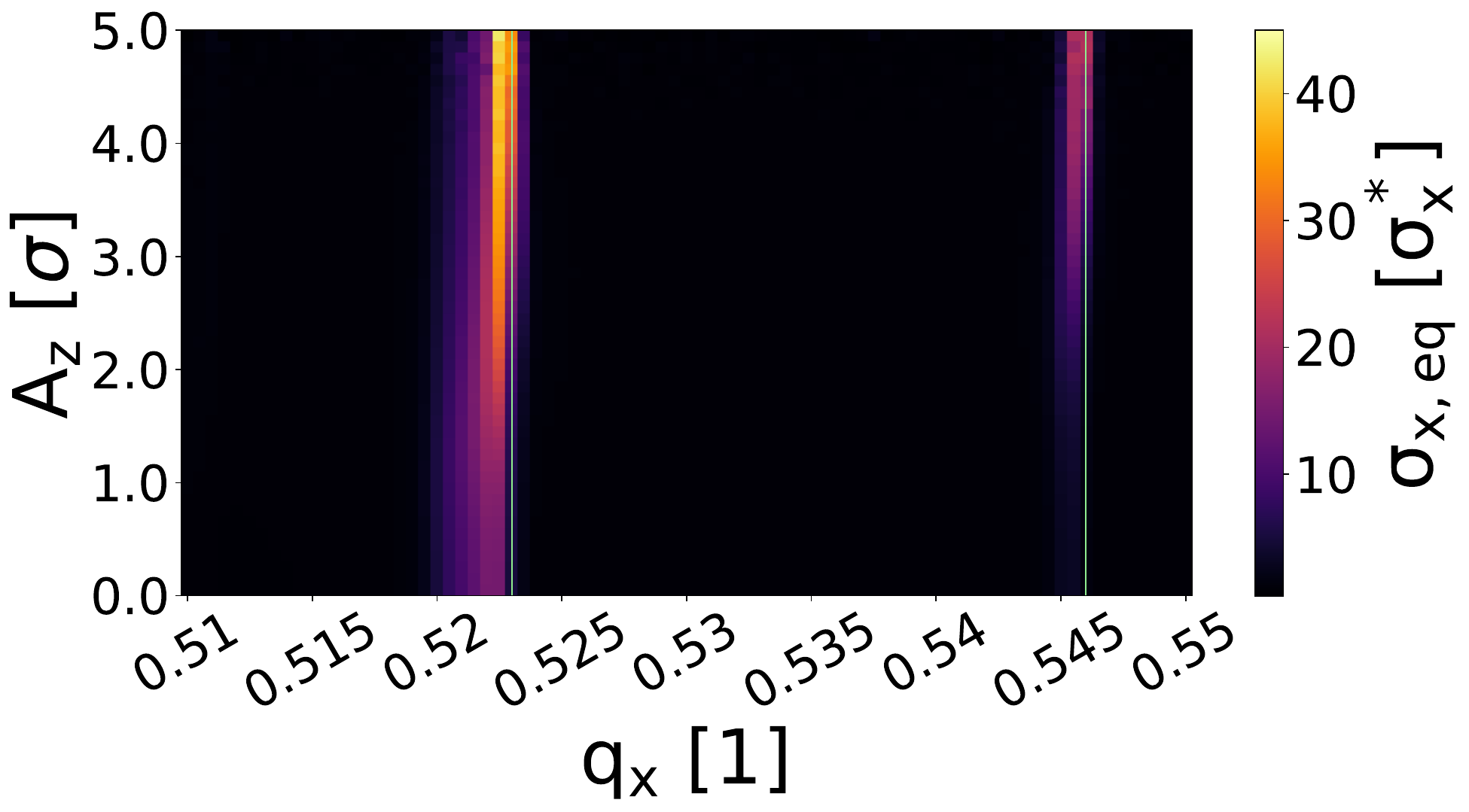}}
	\\	
    \subfloat[\label{fig:sx_b}With longitudinal wakefield.]{
	\includegraphics[width=\columnwidth]{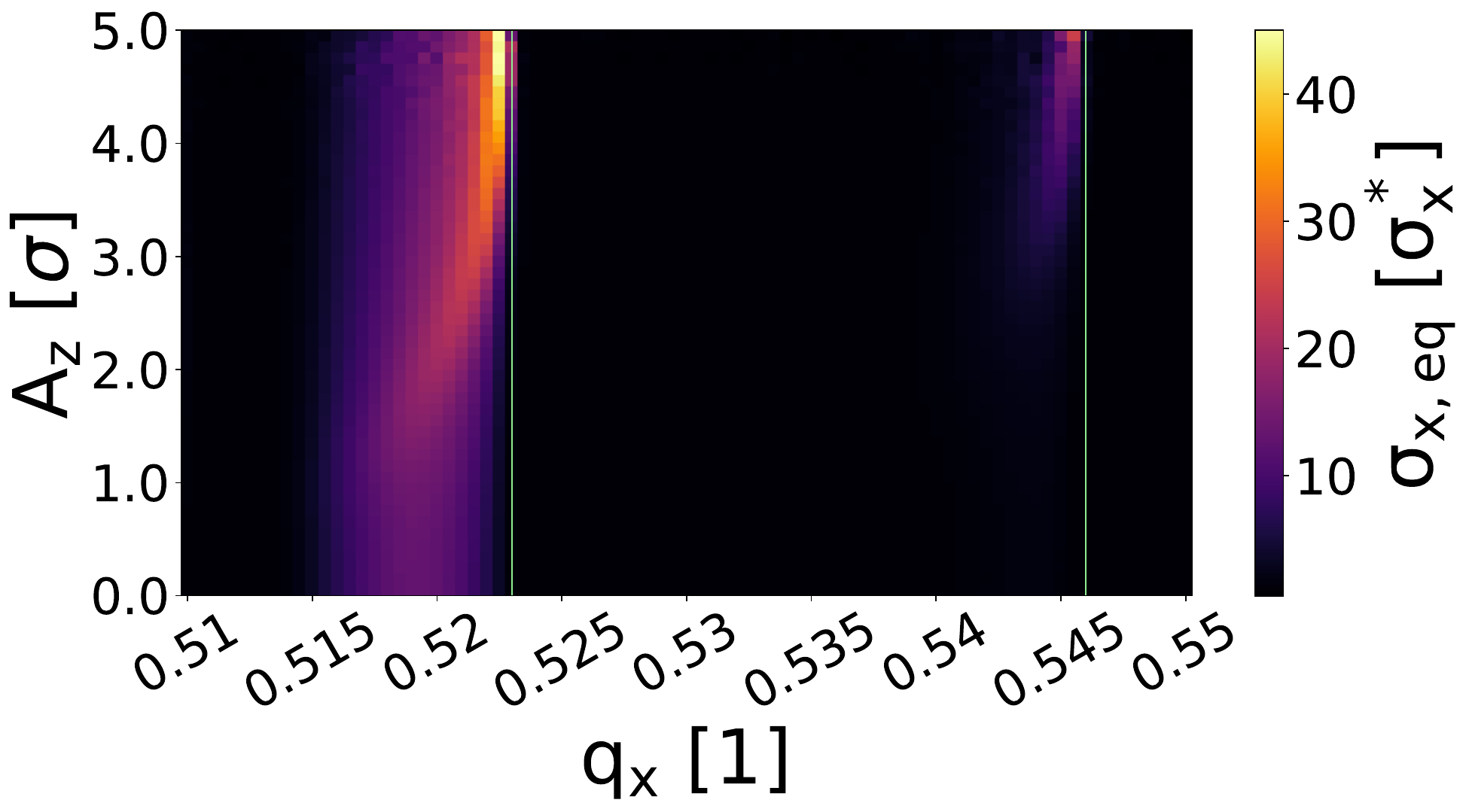}}
    \caption{Horizontal equilibrium RMS beam size simulated with \texttt{Xsuite}, calculated from the last 5000 turns as a function of the horizontal tune $q_x$ and longitudinal amplitude $A_z$ of the particles in the weak bunch, without (top) and in the presence of longitudinal wakefields (bottom). The crab-waist scheme is included in both simulations. The nominal location of the first and second synchrotron sidebands, i.e. $Q_x=0.5+Q_z$ and $Q_x=0.5+2Q_z$, are marked with green vertical lines.}
    \label{fig:sx}
\end{figure}

The presence of resonances affects the particles in the bunch with varying longitudinal actions according to Eqs.~\eqref{eq:Fmx0mz4} and~\eqref{eq:Fmx0mz5}, depending on the tunes according to Eq.~\eqref{eq:resonancexz}. The tracking data used to produce Fig.~\ref{fig:qx_scan_ws} are further analyzed in Fig.~\ref{fig:sx}, which shows the equilibrium horizontal RMS bunch size as a function of the horizontal fractional tune $q_x$. The data are analyzed by breaking it down based on the normalized longitudinal phase space amplitude $A_z$, defined for each particle as:
\begin{equation}
    A_z = \sqrt{\left(\frac{z}{\sigma_z}\right)^2 + \left(\frac{\delta}{\sigma_\delta}\right)^2}.
\end{equation}
Although this definition is an approximation of $A_z$ as given in Eq.~\eqref{eq:Vmx0mz2}, it is particularly useful for studying amplitude-dependent effects in macroparticle tracking simulations.

Figure~\ref{fig:sx} clearly shows that the horizontal blowup strongly depends on the longitudinal amplitude, particularly at horizontal tunes close to the synchrotron sidebands ($2Q_x-m_zQ_z=\textit{Integer}$), roughly following the scaling law of Eq.~\eqref{eq:F2mz1}. In the region where $A_z\gtrsim3$, the blowup becomes more significant and does not decay as expected from Fig.~\ref{fig:F2mz}. This could be explained as follows: when particles that experience synchronized kicks are driven to a large horizontal amplitude $A_x$, they undergo greater perturbations in accordance with Eq.~\eqref{eq:F2mz1}.

The horizontal blowup around the resonance $(m_x,m_z)=(2,4)$ is weaker than that around the resonance $(m_x,m_z)=(2,2)$. This is because the strength of the SBRs decreases as the longitudinal index $m_z$ increases~\cite{dikansky2009effect}, as illustrated in Fig.~\ref{fig:F2mz}. Following this trend, moving the horizontal tune $q_x$ farther away from the half-integer should help mitigate horizontal blowup and its associated detrimental effects~\cite{Shatilov2017ICFABD}.

Figure~\ref{fig:sx_b} shows that the longitudinal wakefield mainly affects particles with relatively small $A_z$. This is because, in this region, the wake kick is more pronounced, leading to a relatively large synchrotron tune shift.

\subsection{Mitigation of incoherent horizontal blowup}

As demonstrated in the earlier sections, the SBRs caused by the interplay between beam-beam interactions and longitudinal wakefields lead to horizontal blowup if the working point is too close to the synchrotron sidebands. In other words, this occurs when the beam’s tune footprint significantly overlaps with wide synchrotron stopbands. Here, we suggest two mitigation techniques for the horizontal blowup. 

\begin{figure}
    \subfloat[\label{fig:qx_scan_bx_ws_x}Horizontal RMS.]{
	\includegraphics[width=\columnwidth]{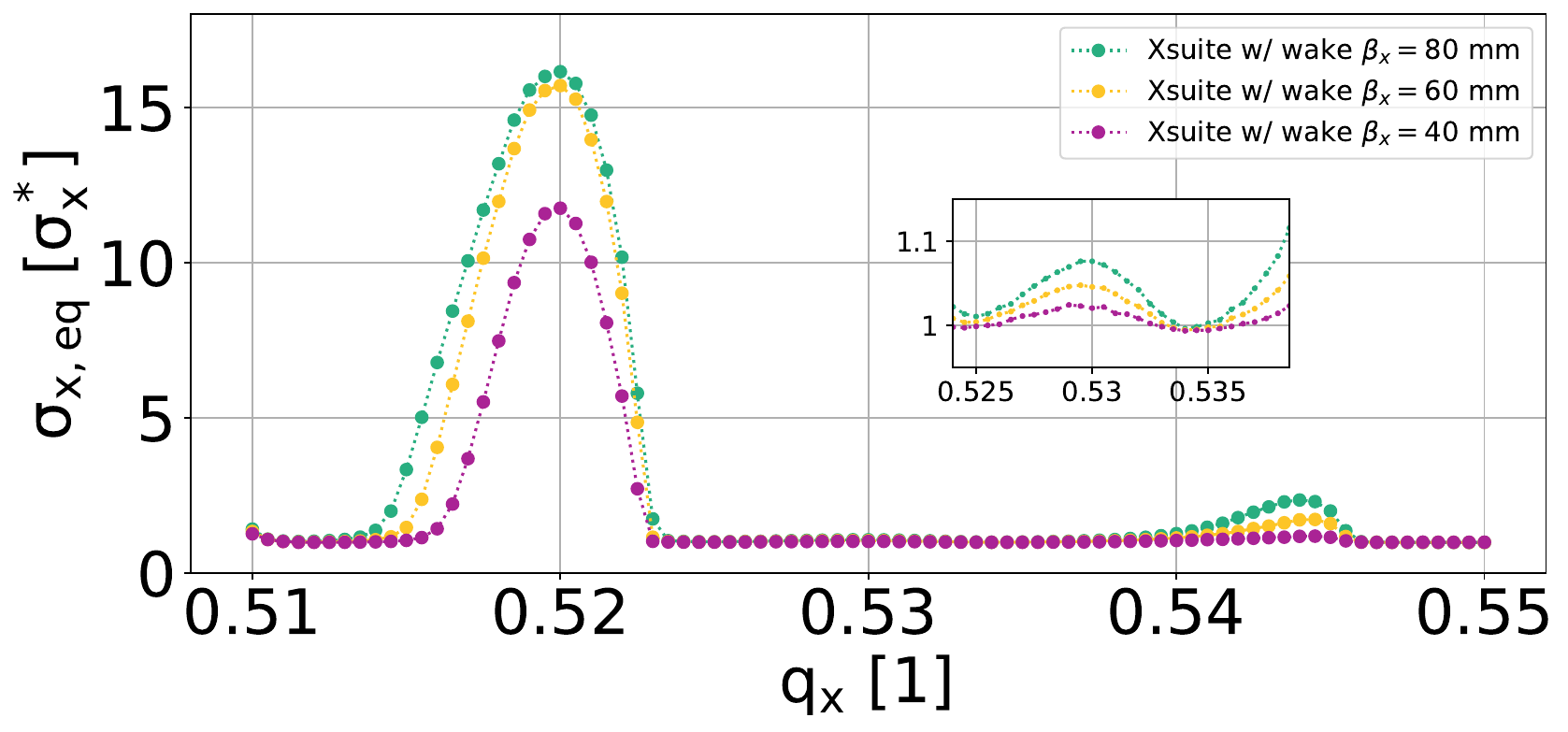}}
	\\	
    \subfloat[\label{fig:qx_scan_bx_ws_y}Vertical RMS.]{
	\includegraphics[width=\columnwidth]{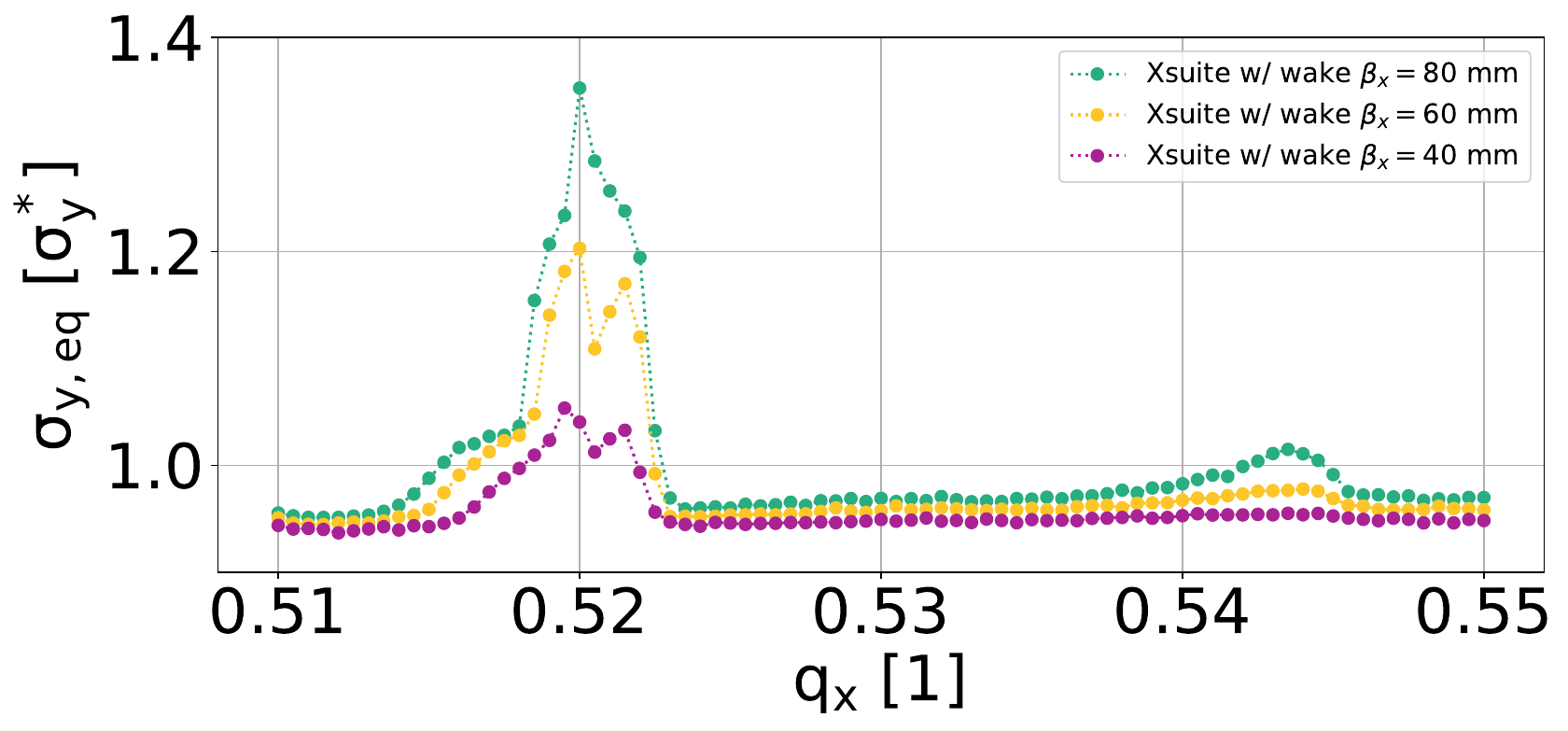}}
	\\	
    \subfloat[\label{fig:qx_scan_bx_ws_lumi}Luminosity.]{
        \includegraphics[width=\columnwidth]{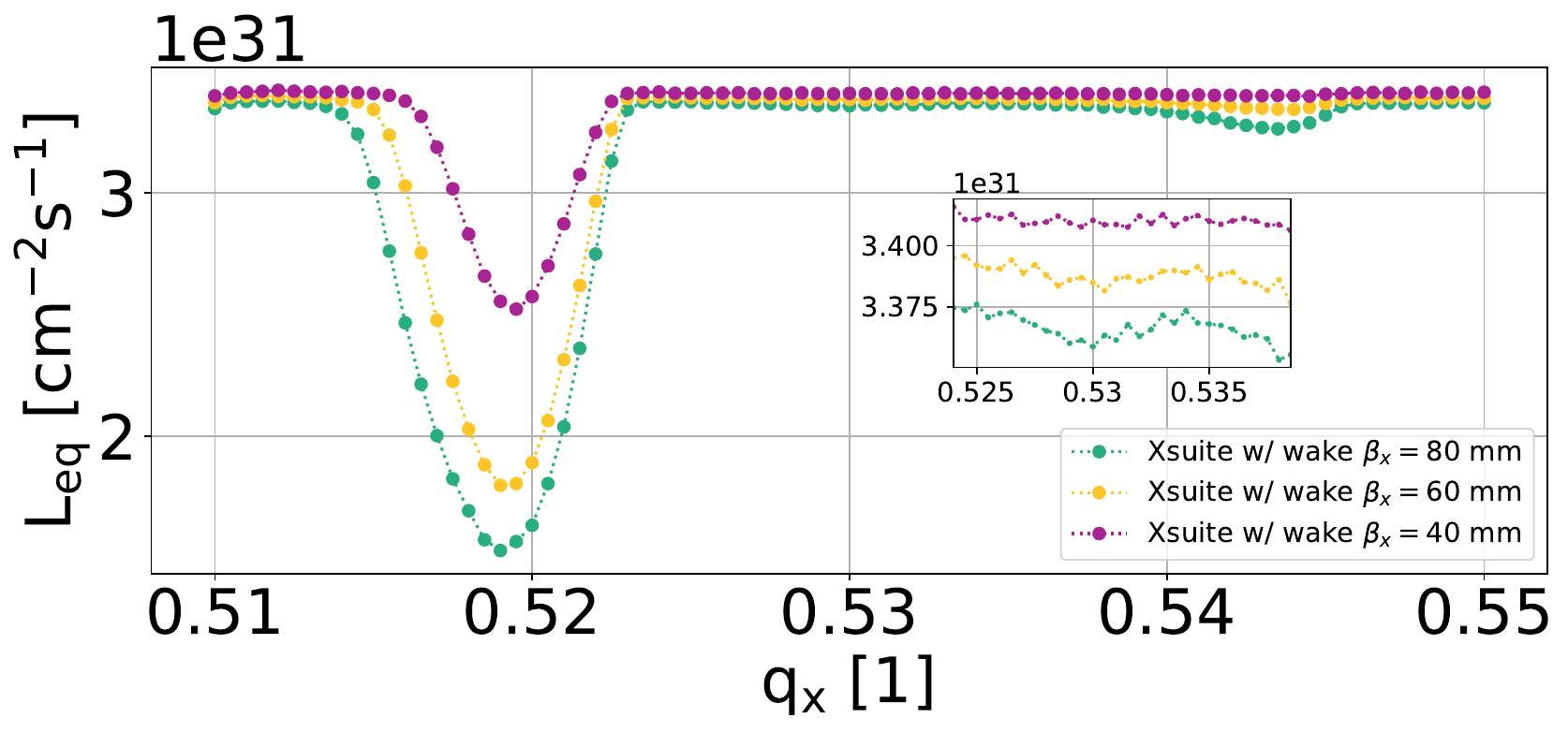}}
    \caption{Equilibrium RMS beam sizes of the LER and luminosity, calculated from the last 5000 turns as a function of the horizontal tune $Q_x$, with wakefields, simulated with \texttt{Xsuite}, with different values of $\beta_x^*$. All simulations include a beam-beam collision in the weak-strong model, including the crab-waist scheme.}
    \label{fig:qx_scan_bx_ws}
\end{figure}

One idea is to decrease $\beta_x^*$. Figure~\ref{fig:qx_scan_bx_ws} show the equilibrium transverse RMS bunch sizes as well as the equilibrium luminosity per bunch crossing as a function of $q_x$, with three different $\beta_x^*$ values for the SuperKEKB LER beam, which we track in our weak-strong model. In these simulations, the HER beam parameters remained unchanged. The inset plot in the top and bottom figures is a zoom into the region around $(m_x,m_z)=(2,3)$, around $q_x=0.53$, according to Eq.~\eqref{eq:Fmx0mz5}, which is only excited when the longitudinal wakefield is included. It can be seen that with the reduction of $\beta_x^*$ by a factor 2 this sideband can be reduced to a negligible level and the luminosity increased to the unperturbed level. Similarly, the maximum amplitude of the other sidebands in both transverse planes is reduced depending on $\beta_x^*$. A reduction of $\beta_x^*$ by such a magnitude is not expected to increase beamstrahlung intensity, as the average beamstrahlung parameter~\cite{Chen} is $\Upsilon_{\text{avg}}\sim10^{-6}$, thus the increase of the equilibrium bunch length can still be considered negligible. For comparison, $\Upsilon_{\text{avg}}\sim10^{-4}$ for the FCC-ee.

It is worth pointing out the effectiveness of this mitigation technique in the vertical blowup, shown in Fig.~\ref{fig:qx_scan_bx_ws_y}. A smaller $\beta_x^*$ for the LER beam indirectly weakens the nonlinear transverse $x-y$ coupling that is caused by the beam-beam interaction, by reducing the horizontal beam-beam tune shift (Eq.~\eqref{eq:xi_x0}) and the horizontal SBR strength (Eq.~\eqref{eq:F2mz1}), thereby the horizontal blowup. This, in turn, enhances the effectiveness of vertical suppression by the crab-waist scheme, which is proportional to $xp_y^2$~\cite{PhysRevAccelBeams.19.111005}, thus mitigating vertical blowup.

The horizontal blowup with a tune around the synchrotron sidebands $2Q_x-m_zQ_z=\textit{Integer}$ with odd $m_z$ could potentially also be mitigated by tuning the phase of the RF system. It can be seen from Eq.~\eqref{eq:Fmx0mz5} that $z_m$ (the peak position of the longitudinal bunch profile), which becomes nonzero due to the wakefield interplay, breaks the symmetry of the integral. This offset in the peak position can be corrected by changing the RF system phase, thereby partially restoring the symmetry and making the $m_z=\textit{odd}$ resonances weaker.

\section{\label{sec:summary}Summary and outlook}

In this paper, we investigated incoherent horizontal SBRs driven by beam-beam interactions in crab-waist colliders. These resonances are modified by the longitudinal wakefields, which induce incoherent synchrotron tune shifts and spreads through potential well distortion. We revisited the theory of these resonances and extended it to include longitudinal wakefields in a simple manner while focusing on the key physics. The crab-waist scheme, designed primarily to suppress betatron resonances, does not significantly influence the dynamics of the horizontal blowup, caused by SBRs. The interplay between beam-beam interactions and longitudinal wakefields broadens the resonant peaks around $2Q_x-m_zQ_z=\textit{Integer}$ with $m_z=2,4,\ldots$. Additionally, due to potential well distortion, a weak horizontal blowup also appears around the resonances $2Q_x-m_zQ_z=\textit{Integer}$ with $m_z=3,5,\ldots$. These effects may influence the selection of horizontal tunes in crab-waist colliders.

As a demonstration, we conducted extensive simulations using various codes to analyze the equilibrium beam dynamics as a function of the horizontal tune and investigated the amplitude dependence of the blowup in the context of the SuperKEKB LER. The simulation results provide valuable insights into horizontal SBRs driven by beam-beam interactions and their interplay with impedance effects. In addition, we have proposed two techniques to mitigate the transverse blowup, which open the door to future studies in this direction. Furthermore, the findings are closely aligned with the theoretical framework outlined in this paper.

\acknowledgements{
P. Kicsiny conducted the \texttt{Xsuite} and \texttt{PyHEADTAIL} simulations and D. Zhou derived the theory in Sec.~\ref{sec:theory} and conducted the simulations with \texttt{BBWS} and the Vlasov solver. The authors would like to thank W. Herr for his useful comments on the manuscript. This work was carried out under the auspices of and with support from the Swiss Accelerator Research and Technology (CHART) programme (\url{www.chart.ch} and with funding from the Europe-America-Japan Accelerator Development and Exchange Programme (EAJADE). EAJADE is a Marie Sklodowska-Curie Research and Innovation Staff Exchange (SE) action, funded by the EU under the Horizon-Europe Grant agreement ID: 101086276 (\url{https://cordis.europa.eu/project/id/101086276}).}

\appendix
\bibliography{apssamp} 
\end{document}